\documentclass[usenatbib]{mnras}
\usepackage{epsfig}
\usepackage{psfig}
\usepackage{amsmath,amssymb}
\usepackage{txfonts}
\usepackage[english]{babel}
\usepackage[utf8]{inputenc}
\usepackage[T1]{fontenc}
\usepackage{fontenc}
\usepackage{soul}
\allowdisplaybreaks
\def\los/{{line-of-sight}}
\def\Los/{{Line-of-sight}}
\def\av#1{\left\langle#1\right\rangle}

\newcommand\Mstar{M_{\star}}
\newcommand\Mstarmain{M_{\star,{\rm main}}}
\newcommand\Mstarremn{M_{\star,{\rm remn}}}
\newcommand\Mstarsat{M_{\star,{\rm sat}}}
\newcommand\MDM{M_{\rm DM}}
\newcommand\MDMzero{M_{\rm DM,0}}
\newcommand\rstar{r_{\star}}
\newcommand\rstarmain{r_{\star,{\rm main}}}
\renewcommand\Re{R_{\rm e}}
\newcommand\Remain{R_{\rm e,main}}
\newcommand\Resat{R_{\rm e,sat}}
\newcommand\Resattilde{\tilde{R}_{\rm e,sat}}
\newcommand\Reremntilde{\tilde{R}_{\rm e,remn}}
\newcommand\Reremn{R_{\rm e,remn}}
\newcommand\AIMF{A_{\rm IMF}}
\newcommand\BIMF{B_{\rm IMF}}
\newcommand\CIMF{C_{\rm IMF}}

\newcommand\xistar{\xi_{\star}}

\newcommand\avxistar{\av{\xi}_{\star}}
\newcommand\alphaIMF{\alpha_{\rm IMF}}
\newcommand\alphaIMFi{\alpha_{{\rm IMF},i}}
\newcommand\alphae{\alpha_{\rm e}}
\newcommand\alphaemain{\alpha_{\rm e,main}}
\newcommand\alphaesat{\alpha_{\rm e,sat}}
\newcommand\alphaIMFmain{\alpha_{\rm IMF,main}}
\newcommand\alphaIMFsat{\alpha_{\rm IMF,sat}}
\newcommand\alphaesattilde{\tilde{\alpha}_{\rm e,sat}}
\newcommand\alphaeremn{\alpha_{\rm e,remn}}
\newcommand\alphaeremntilde{\tilde\alpha_{\rm e,remn}}
\newcommand\xlos{x_{\rm los}}

\newcommand\conc{c}
\renewcommand\d{{\rm d}}

\newcommand\E{\mathcal{E}}

\newcommand\Eq{equation~}

\newcommand\Fig{Fig.~}
\newcommand\Figs{Figs.~}
\newcommand\gammasigma{{\gamma_\sigma}}
\newcommand\gammaalpha{{\gamma_\alpha}}

\newcommand\lu{{\ell_{\rm u}}}

\newcommand\kms{{\rm \,km\,s^{-1}}}
\newcommand\kpc{{\rm \,kpc}}

\newcommand\M{M}
\newcommand\Mu{{\mathcal M}_{\rm u}}

\newcommand\Msun{{\M}_{\odot}}

\newcommand\Psitotzero{\Psi_{\rm tot,0}}
\newcommand\Qtilde{\tilde{Q}}

\newcommand\rhostar{\rho_{\star}}

\newcommand\rs{r_{\rm s}}
\newcommand\rstilde{\tilde{r}_{\rm s}}
\newcommand\rperi{r_{\rm peri}}

\newcommand\ra{r_{\rm a}}

\newcommand\rhoDM{\rho_{\rm DM}}
\newcommand\PsiDM{\Psi_{\rm DM}}
\newcommand\Psitot{\Psi_{\rm tot}}
\newcommand\Psistar{\Psi_\star}

\newcommand\rvir{{r_{\rm vir}}}

\newcommand\sigmae{\sigma_{\rm e}}
\newcommand\sigmaemain{\sigma_{\rm e,main}}
\newcommand\sigmaesat{\sigma_{\rm e,sat}}
\newcommand\sigmaeremn{\sigma_{\rm e,remn}}
\newcommand\sigmaesattilde{\tilde\sigma_{\rm e,sat}}
\newcommand\sigmaeremntilde{\tilde\sigma_{\rm e,remn}}
\newcommand\sigmalos{\sigma_{\rm los}}
\newcommand\Sect{Section~}
\newcommand\Sects{Sections~}

\newcommand\Tab{Table~}

\newcommand\vv{{\boldsymbol v}}

\newcommand{\vvRv}{{\boldsymbol v}_{\boldsymbol R}}
\newcommand\vlos{v_{\rm los}}
\newcommand\vlosi{v_{{\rm los},i}}
\newcommand\vbarlos{\overline{v}_{\rm los}}
\newcommand\vbarlosk{\overline{v}_{{\rm los},k}}

\begin{document}

\date{Accepted, September 11 2020}

\title[Velocity dispersion and IMF in galaxy mergers]{Stellar velocity dispersion and initial mass function gradients in dissipationless galaxy mergers} 

\author[C.\ Nipoti et al.]{\parbox{\textwidth}{Carlo Nipoti$^{1}$\thanks{E-mail: carlo.nipoti@unibo.it}, Carlo  Cannarozzo$^{1,2}$, Francesco Calura$^{2}$, Alessandro Sonnenfeld$^{3}$ and
    Tommaso Treu$^{4}$}\vspace{0.4cm}\\
  \parbox{\textwidth}{
$^1$Department of Physics and Astronomy, University of Bologna,
  via Gobetti 93/2, I-40129 Bologna, Italy \\
  $^2$INAF - Astrophysics and Space Science Observatory of Bologna, via Gobetti 93/3, I-40129, Bologna, Italy \\
  $^{3}$Leiden Observatory, Leiden University, Niels Bohrweg 2, 2333 CA Leiden, The Netherlands\\
$^{4}$Department of Physics and Astronomy, University of California, Los Angeles, CA, 90095-1547, USA\\  
}}

%

\maketitle 

\begin{abstract}
The stellar initial mass function (IMF) is believed to be
non-universal among early-type galaxies (ETGs). Parameterizing the IMF
with the so-called IMF mismatch parameter $\alphaIMF$, which is a
measure of the stellar mass-to-light ratio of an ensemble of stars and
thus of the `heaviness' of its IMF, one finds that for ETGs $\alphae$
(i.e.\ $\alphaIMF$ integrated within the effective radius $\Re$)
increases with $\sigmae$ (the line-of-sight velocity dispersion
$\sigmalos$ integrated within $\Re$) and that, within the same ETG,
$\alphaIMF$ tends to decrease outwards.  We study the effect of
dissipationless (dry) mergers on the distribution of the IMF mismatch
parameter $\alphaIMF$ in ETGs using the results of binary major and
minor merging simulations. We find that dry mergers tend to make the
$\alphaIMF$ profiles of ETGs shallower, but do not alter significantly
the shape of the distributions in the spatially resolved
$\sigmalos\alphaIMF$ space. Individual galaxies undergoing dry mergers
tend to decrease their $\alphae$, due to erosion of $\alphaIMF$
gradients and mixing with stellar populations with lighter IMF. Their
$\sigmae$ can either decrease or increase, depending on the merging
orbital parameters and mass ratio, but tends to decrease for
cosmologically motivated merging histories. The $\alphae$-$\sigmae$
relation can vary with redshift as a consequence of the evolution of
individual ETGs: based on a simple dry-merging model, ETGs of given
$\sigmae$ are expected to have higher $\alphae$ at higher redshift,
unless the accreted satellites are so diffuse that they contribute
negligibly to the inner stellar distribution of the merger remnant.
\end{abstract}


\section{Introduction}
\label{sec:intro} 

There is growing evidence that the stellar initial mass function (IMF)
in not universal.  Not only do different galaxies have different
stellar IMFs, but, at least in some cases, there are indications that
the IMF is also different in different regions of the same galaxy
\citep[see e.g.][]{Hop18,Smi20}.  When the single stars are not
resolved in observations, we have access only to indirect information
about the stellar IMF, such as, for instance, the so called IMF
mismatch parameter\footnote{In the literature the IMF mismatch
  parameter $\alphaIMF$ is sometimes referred to as effective IMF,
  mass excess or excess stellar mass-to-light ratio.}
\begin{equation}
\alphaIMF\equiv\frac{(\Mstar/L)_{\rm true}}{(\Mstar/L)_{\rm ref}},
\label{eq:alphaimf}
\end{equation}  
 where $(\Mstar/L)_{\rm true}$ is the true stellar mass-to-light ratio
 of an ensemble of stars (in a given band) and $(\Mstar/L)_{\rm ref}$
 is the stellar mass-to-light ratio (in the same band) that one would
 infer assuming a reference IMF, for instance the \citet{Sal55}, the
 \citet{Kro01} or the \citet{Cha03} IMF \citep{Tre10}. $\alphaIMF$ is
 a quantity integrated over the stellar population, which, per se,
 does not contain information on the {\em shape} of the IMF.  Broadly
 speaking, IMFs with high $\alphaIMF$ are said to be `heavy': a high
 value of $\alphaIMF$ can be due to an overabundance of low-mass stars
 (bottom-heavy IMF), but also of high-mass stars (top-heavy IMF;
 \citealt{Bas10}).

When large samples of massive early-type galaxies (ETGs) are
considered, an empirical correlation is found between the IMF and the
central stellar velocity dispersion, in the sense that ETGs with
higher velocity dispersion have, on average, heavier IMF
\citep[][]{Tre10,Cap12,Con12,Dut12,Tor13,Spi14,Li17,Ros18}.  For
instance, for a sample of present-day ETGs with $90\kms\lesssim
\sigmae\lesssim 270\kms$, taking as reference IMF the Salpeter IMF,
\citet{Pos15} find
\begin{equation}
  \log\alphae=(0.38\pm0.04)\log\left(\frac{\sigmae}{200\kms}\right) +(-0.06\pm0.01),
\label{eq:obsalpsig}  
\end{equation}  
with intrinsic scatter $0.12$ dex in $\alphae$ at given $\sigmae$, where
the {\em effective IMF mismatch parameter} $\alphae$ is the IMF
mismatch parameter measured within the effective radius $\Re$ and the
{\em effective velocity dispersion} $\sigmae$ is the central stellar
velocity dispersion measured within $\Re$.  It must be stressed that
some individual galaxies are found to deviate substantially from this
relation: in particular, there are cases of massive (high-$\sigmae$)
ETGs with light IMF \citep{Smi15,Col18,Son19}.

Spatially resolved estimates of $\alphaIMF$ in ETGs have revealed the
presence of IMF gradients: $\alphaIMF$ is higher in the centre than in
the outer regions of individual galaxies, ranging from an IMF heavier
than the Salpeter IMF to a lighter IMF with Chabrier-like $\alphaIMF$
(\citealt{Mar15a,Dav17,vDo17,Zie17,Old18,Son18,LaB19}; but see also
\citealt{Alt17,Alt18} and \citealt{Vau18}, who find no significant IMF
radial gradients in their samples of ETGs).

In the currently favoured hierarchical model of galaxy formation, ETGs
are believed to form in two phases \citep{Ose10}: a first mostly
dissipative phase of in-situ star formation at $z\gtrsim 2$ and a
second phase of accretion of stars mainly via dissipationless (`dry')
mergers at $z\lesssim 2$ (see also section 10.8 of \citealt*{CFN19}).
In this context, the correlation between $\alphae$ and $\sigmae$
observed for present-day massive ETGs must be produced by the
combination of the above two phases. Focusing on the second phase,
\citet*{Son17} studied the redshift evolution of the
$\alphae$-$\sigmae$ relation since $z\approx 2$, considering
cosmologically motivated merging hierarchies in the simple case in
which all the mergers are dissipationless and the stellar populations
mix completely in the mergers.  \citet{Son17} found that, as a
consequence of the accretion of lower-mass satellites, both $\sigmae$
and $\alphae$ of massive ETGs decrease with cosmic time. Nevertheless,
in this model the $\alphae$-$\sigmae$ relation remains essentially
unaltered as cosmic time goes on, because individual massive ETGs move
in the $\sigmae\alphae$ plane roughly along the $\alphae$-$\sigmae$
relation.

The model of \citet{Son17} is based on a few simplifying assumptions,
which are only partially justified: the stellar velocity dispersion is
assumed to be proportional to the host halo virial velocity
dispersion, neither dissipation nor star formation is allowed during
the merger, and any gradient in the $\alphaIMF$ distribution within
galaxies is neglected. \citet*{Bla17} studied the same process with a
more realistic, though not fully self-consistent approach, performing
a post-processing analysis of the Illustris cosmological hydrodynamic
simulation \citep{Vog14}. \citet{Bla17} found that, at the time of
star formation, $\alphaIMF$ must vary strongly with the local velocity
dispersion in order to reproduce the observed $\alphae$-$\sigmae$
relation of present-day ETGs.  More recently
\citet{Bar18,Bar19a,Bar19b} revisited the problem of the theoretical
origin and evolution of the $\alphae$-$\sigmae$ relation of ETGs in a
more self-consistent way by performing cosmological hydrodynamic
simulations in which, when stars form, the IMF depends on the local
pressure of the gas.  The models considered by Barber at al., which
are calibrated to reproduce the observed present-day trend of
$\alphae$ increasing with $\sigmae$, assume that in higher-pressure
environments the stars form with heavier IMF, either bottom-heavy
(model `LoM') or top-heavy (model `HiM'). \citet{Bar19b} find that, at
given $\sigmae$, the average $\alphae$ of ETGs tends to be higher at
higher $z$ for model `LoM', in which $\alphaIMF$ is essentially
independent of the age of the stellar population, and lower at higher
$z$ for model `HiM', in which instead $\alphaIMF$ is significantly
lower for younger stellar populations. The question of the
non-universality of the IMF in the context of galaxy formation and
evolution has been explored also with semi-analytical models in
various papers \citep{Nag++05, Cal09, Cha++15, GargiuloI++15, Fon17},
which however focus on the evolution of the chemical properties of
galaxies without exploring specifically correlations of the IMF with
the stellar velocity dispersion.

In this paper, we approach theoretically the question of the evolution
of the $\alphaIMF$ of ETGs by taking into account in detail
$\alphaIMF$ gradients within galaxies. In a cosmological context, the
$\alphaIMF$ gradients of simulated ETGs have been analyzed by both
\citet{Bla17} and \citet{Bar19b}, who find that their simulated
present-day ETGs have broadly realistic $\alphaIMF$ profiles. In the
case of \citet{Bar19b}, the $\alphaIMF$ profiles tend to be steeper
for model `LoM' than for model `HiM'. Here we address the question of
the evolution of the $\alphaIMF$ gradients with a simpler approach,
using binary dissipationless merging simulations, which are outside a
fully cosmological framework. Though idealized in some respect, our
simulations allow us to study in great detail the dynamical effects on
the distribution of $\alphaIMF$, which are believed to be important in
the second phase of ETG formation.  With our approach, we can
disentangle these dynamical effects from other effects more directly
related to dissipation and star formation, which is not
straightforward in cosmological hydrodynamic simulations.

If the IMF is not universal, the presence of IMF gradients within ETGs
must be expected, because merging produces a partial mixing of the
stellar populations \citep{Whi80}.  For instance, in the idealized
case of a binary dissipationless merger in which the two merging
galaxies have different $\alphaIMF$ and no $\alphaIMF$ gradient, an
$\alphaIMF$ gradient will naturally arise in the remnant. More
realistically, we can envisage an evolutionary scenario for ETGs in
which galaxies with $\alphaIMF$ gradients merge with other galaxies
that have themselves $\alphaIMF$ gradients and generally different
average $\alphaIMF$. This is the approach we adopt in the models here
presented.

The paper is organized as follows.  The set-up and the analysis of the
$N$-body simulations are described in \Sect\ref{sec:simu} and
\Sect\ref{sec:analysis}, respectively. Our results are presented in
\Sect\ref{sec:res}. \Sect\ref{sec:concl} concludes.

\section{$N$-body simulations} 
\label{sec:simu} 

\subsection{Sets of simulations}
\label{sec:set}

\begin{table*}
\caption{Parameters of the galaxy encounters, of the progenitor galaxy
  models and of the merger remnants in the dissipationless
  binary-merger $N$-body simulations analyzed in this work. Name: name
  of simulation. Set: name of the simulation set.  $\MDM$: total DM
  mass. $M_\star$: total stellar mass. $\conc$: NFW
  concentration. $r_s$: NFW scale radius. $\Re$: effective
  radius. $\xistar=\Mstarsat/\Mstarmain$: stellar mass
  ratio. $\rperi/\rvir$: pericentric-to-virial radius ratio ($\rvir$
  is the virial radius of the main galaxy), quantifying the orbital
  angular momentum of the encounter (all encounters are
  parabolic). $\AIMF$, $\BIMF$ and $\CIMF$: parameters used in
  post-processing to assign $\alphaIMF$ to particles
  (equations~\ref{eq:alphaq1} and ~\ref{eq:alphaq2}). Here
  $\Resattilde\equiv\Resat/\Remain$,
  $\sigmaesattilde\equiv\sigmaesat/\sigmaemain$,
  $\alphaesattilde\equiv\alphaesat/\alphaemain$,
  $\Reremntilde\equiv\Reremn/\Remain$,
  $\sigmaeremntilde\equiv\sigmaeremn/\sigmaemain$ and
  $\alphaeremntilde\equiv\alphaeremn/\alphaemain$.  In all cases
  {$(\MDM/\Mstar)_{\rm main}=49$}, {$\conc_{\rm main}=8$} and
  {$(r_s/\Re)_{\rm main}=11.6$}.  Subscripts `main', `sat' and `remn'
  refer to the main galaxy, the satellite galaxy and the remnant,
  respectively.  }
\label{tab:simu} 
\begin{tabular}{llcccccccccccccc} 
{Name} 
&{Set}   
& {$(\MDM/M_\star)_{\rm sat}$}   
& {$\conc_{\rm sat}$}   
& {$(r_s/\Re)_{\rm sat}$} 
  & {$\xi_\star$}
& {$\rperi/\rvir$}
& {$\AIMF$}
& {$\BIMF$} 
& {$\CIMF$}
& $\Resattilde$
& $\Reremntilde$
& $\sigmaesattilde$
& $\sigmaeremntilde$
& $\alphaesattilde$
& $\alphaeremntilde$
\\
\hline
1Dh         &D    & 49 & 8  & 11.6 & 1 & 0 & 2 & 0 & 1 &1 & 2.21 &1 & 1.030 &1 & 0.980 \\ %
1Do         &D    & 49 & 8  & 11.6 & 1 & 0.12 & 2 & 0 & 1 & 1& 1.93 &1 &1.080 & 1& 0.983 \\ %
1Dh\_bis    &D    & 49 & 8  & 11.6 & 1 & 0 & 0.5 & 1 & 1 &1& 2.21  &1 &1.030 &1 & 0.995\\ %
1Do\_bis    &D    & 49 & 8  & 11.6 & 1 & 0.12 & 0.5 & 1 & 1 & 1& 1.93 &1 &1.080 &1 & 0.996\\ %
0.5Dh       &D    & 49 & 8  & 11.6 & 0.5 & 0 & 2 & 0 & 0.95      & 0.66 & 1.96 & 0.870 & 0.977 &0.950 &0.962\\ %
0.5Do       &D    & 49 & 8  & 11.6 & 0.5 & 0.2 & 2 & 0 & 0.95    & 0.66 & 1.59 & 0.870 & 1.036&0.950 &0.966\\ %
0.2Dh       &D    & 49 & 8  & 11.6 & 0.2 & 0 & 2 & 0 & 0.895     & 0.38 & 1.65 & 0.746 & 0.917 &0.895 &0.971 \\ %
0.2Do       &D    & 49 & 8  & 11.6 & 0.2 & 0.2 & 2 & 0 & 0.895   & 0.38 & 1.38 & 0.746 &0.983 &0.895 &0.973\\ %
0.2D3h      &D3   & 35 & 8.5  & 8.8  & 0.2 & 0 & 2 & 0 & 0.935   & 0.38 & 1.73 & 0.758 &0.906 &0.903 &0.970\\ %
0.2D3o      &D3   & 35 & 8.5  & 8.8  & 0.2 & 0.2 & 2 & 0 & 0.935 & 0.38 & 1.40 & 0.758 &0.981 &0.903 &0.972\\ %
0.2D4h      &D4   & 75 & 8.5  & 15.0 & 0.2 & 0 & 2 & 0 & 0.85    & 0.38 & 1.61 & 0.739& 0.917&0.892 &0.973\\ %
0.2D4o      &D4   & 75 & 8.5  & 15.0 & 0.2 & 0.2 & 2 & 0 & 0.85  & 0.38 & 1.36 & 0.739&0.980 &0.892 &0.973\\ %
\hline 
\end{tabular} 
\end{table*}

In this paper we analyze $N$-body simulations of binary
dissipationless galaxy mergers presented in previous works.  In
particular, we focus on ten simulations taken from the simulation sets
named D, D3 and D4 in \citet*{Son14}, some of which were originally
presented in \citet{NTB09} and \citet{Nip++12}.

The six simulations of set D are characterized by different values of
the stellar mass ratio $\xistar=\Mstarsat/\Mstarmain$ between the two
progenitor galaxies: two with $\xistar=1$ (equal-mass major merger), two
with $\xistar=0.5$ (unequal-mass major merger) and two with
$\xistar=0.2$ (minor\footnote{Mergers are usually classified as minor
  when the mass ratio is lower than either $1/3$ or $1/4$.} merger).
Here $\Mstarmain$ is the stellar mass of the more massive progenitor galaxy
(hereafter referred to as {\em main} galaxy) and $\Mstarsat\leq
\Mstarmain$ is the stellar mass of the less massive progenitor galaxy
(hereafter referred to as {\em satellite} galaxy). Throughout the
paper the subscripts `main', `sat' and `remn' indicate quantities
relative to the main galaxy, to the satellite galaxy, and
to the merger remnant, respectively.

The main galaxy is modelled as a two-component spherically symmetric
stellar system with a stellar component and a dark matter (DM)
halo. This model is described in \Sect\ref{sec:progenitor}.  In the
case of equal-mass mergers ($\xistar=1$) the satellite galaxy is
identical to the main galaxy. In the case of unequal-mass mergers
($\xistar<1$) of set D, the satellite galaxy is a smaller-scale
replica of the main galaxy, i.e.\ it is structurally and kinematically
homologous (see e.g.\ section 5.4 of \citealt{CFN19}) to the main
galaxy, but has different mass and length scale, depending on
$\xistar$ (see \Sect\ref{sec:progenitor}). In the simulations of
set D3 and D4 (all with $\xistar=0.2$) the satellite galaxy is not
homologous to the main galaxy (see \Sect\ref{sec:progenitor}).

In all the simulations considered here the mergers are parabolic
(i.e.\ with zero orbital energy in the point-mass two-body
approximation of the encounter). For each value of $\xistar$ and each
set (D, D3 and D4), we have two simulations, one with zero orbital
angular momentum ($\rperi=0$; hereafter referred to as {\em head-on}
merger simulations) and the other with non-zero orbital angular
momentum ($\rperi\neq0$; hereafter referred to as {\em off-axis}
merger simulations), where $\rperi$ is the pericentric radius. In
Table~\ref{tab:simu}, where the main parameters of the simulations are
reported, the orbital angular momentum is quantified by the ratio
$\rperi/\rvir$, where $\rvir$ is the virial radius of the main galaxy.
We note that the values of $\rperi/\rvir$ adopted for our off-axis
merger simulations are close to the median values found for halo-halo
mergers in cosmological simulations \citep[e.g.][]{Wet11}, which
suggests that the off-axis simulations could be more representative of
cosmologically motivated mergers than the head-on simulations.  For
further details on the set-up of the initial condition we refer the
reader to \citet{Son14} and previous papers \citep{NTB09,Nip++12} from
which some simulations were collected.
  
\subsection{Progenitor galaxy models}
\label{sec:progenitor}

The stellar
density distribution of the progenitor galaxies is represented by a $\gamma$
model \citep{Deh93,Tre94} with $\gamma=3/2$:
\begin{equation}
\rhostar (r)= \frac{3}{8\pi}{\Mstar\rstar \over r^{3/2} (r+\rstar)^{5/2}},
\label{eq:rhostar}
\end{equation}
%
where $\Mstar$ is the total stellar mass and $\rstar$ is the characteristic
radius of the stellar component.   The DM halo
is described by a \citet*[][NFW]{Nav96} model, so the DM density
distribution is
\begin{equation}
\rhoDM (r)={\MDMzero \over r(r+\rs)^2}\exp\left[{-\left({r\over \rvir}\right)^2}\right],
\label{eqrhodm}
\end{equation}
where $\rs$ is the scale radius, $\MDMzero$ is a reference mass and we
adopt an exponential cut-off to truncate the distribution smoothly at
the virial radius $\rvir$, so the  total DM mass 
$\MDM=4\pi \int_0^{\infty}\rhoDM(r)r^2 dr$
is finite.  We assume Osipkov-Merritt \citep[][]{Osi79,Mer85}
anisotropy in the velocity distribution of the stellar component,
whose distribution function is then given by
\begin{equation}
f(Q)=\frac{1}{\sqrt{8}\pi^2}\frac{\d}{\d Q}
       \int_0^Q{\frac{\d{\varrho_\star}}{\d\Psitot}}{\frac{\d\Psitot}{\sqrt{Q-\Psitot}}},
\label{eqdf}
\end{equation}
where
\begin{equation}
\varrho_\star (r)=\left(1+\frac{r^2}{\ra^2}\right)\rhostar (r).
\end{equation}
The variable $Q$ is defined as $Q\equiv \E-{L^2/2\ra^2}$, where the
relative (positive) energy is given by $\E =\Psitot-v^2/2$, $v$ is the
modulus of the velocity vector, the relative (positive) total
potential is $\Psitot=\Psistar +\PsiDM$ ($\Psistar$ and $\PsiDM$ are,
respectively, the relative potentials of the stellar and DM
components), $L$ is the angular momentum modulus per unit mass, and
$f(Q)=0$ for $Q\leq0$. The quantity $\ra$ is the so--called anisotropy
radius: for $r \gg\ra$ the velocity dispersion tensor is radially
anisotropic, while for $r \ll \ra $ the tensor is nearly isotropic.
In the limit $\ra \to\infty$, $Q=\E$ and the velocity dispersion
tensor is globally isotropic.


The orbital distribution of the DM halo is assumed isotropic, so the
distribution function of the DM component is given by
equation~(\ref{eqdf}) where $Q=\E$ and $\rhoDM(r)$ substitutes
$\varrho_\star(r)$.  These $\gamma=3/2$ plus NFW models have four free
parameters: the concentration $\conc\equiv \rvir/\rs$, the
dark-to-stellar mass ratio $\MDM/\Mstar$, the ratio
$\rstilde\equiv\rs/\Re$ and the anisotropy radius $\ra$.  We consider
different choices of these parameters for the progenitor galaxies of
our mergers. When building the progenitor galaxies (both main and
satellite) of set D we assume $\conc=8$, $\MDM/\Mstar=49$,
$\rstilde=11.6$ and $\ra/\rstar=1$. In set D3 and D4 the main galaxy
is identical to that of set D, but the satellite has different values
of the parameters: $\conc=8.5$, $\MDM/\Mstar=35$, $\rstilde=8.8$ and
$\ra/\rstar=1$ for set D3, and $\conc=8.5$, $\MDM/\Mstar=75$,
$\rstilde=15$ and $\ra/\rstar=1$ for set D4. The values of the
parameters are such that the galaxy models are realistic for massive
ETGs \citep{NTB09,Son14}. In all the runs $\Resat/\Remain=\xistar^a$
and $\sigmaesat/\sigmaemain=\xistar^b$ with $a\simeq 0.6$ and $b\simeq
0.2$, so the satellites and the main galaxies lie on $\Re$-$\Mstar$
and $\sigmae$-$\Mstar$ relations with slopes similar to those observed
for massive ETGs \citep[e.g.][]{Cim++12,vdWel14,Can20}.

\subsection{Parameters of  the $N$-body simulations}
\label{sec:para}

All the binary-merger $N$-body simulations were run with the
collisionless $N$-body code {\sc fvfps} \citep[Fortran Version of a
  Fast Poisson Solver;][]{LNC03,NLC03}.  Stellar and dark matter
particles have the same mass in runs with $\xistar=1$; when
$\xistar<1$ the dark matter particles are twice as massive as the
stellar particles.  The total number of particles used in each
simulation is in the range $1.6\times10^6-3.1\times10^6$.  The
parameters of the simulations are given in \citet{NTB09},
\citet{Nip++12} and \citet{Son14}. In all the simulations the galaxy
encounter is followed up to the virialisation of the resulting stellar
system. We define the merger remnant as the system composed by the
bound stellar and dark matter particles at the end of the simulation.

\section{Analysis of the simulations}
\label{sec:analysis}

\subsection{Assigning $\alphaIMF$ to stellar particles}
\label{sec:assignalpha}

Given that the distribution function of the stellar component of each
progenitor galaxy depends on the integral of motion $Q$, we can build
a stationary galaxy model with a gradient in a stellar population
property, say metallicity $Z$, by assigning to each particle a value
of $Z$ as a function of $Q$ (\citealt{Cio95}; see also
\citealt{Nip03b}).  Here we are interested in $\alphaIMF$ gradients,
thus we assign a value of $\alphaIMF$ to each particle in each
progenitor galaxy as a function of $Q$.  Let us consider a binary
merger between the main galaxy of stellar mass $\Mstarmain$ and the
satellite of stellar mass $\Mstarsat\leq \Mstarmain$.

In order to assign a value of $\alphaIMF$ to particles belonging to the main galaxy, we  adopt the linear relation
\begin{equation}
\alphaIMF(Q)=\AIMF\Qtilde+\BIMF,
\label{eq:alphaq1}
\end{equation}
where $\Qtilde\equiv Q/\Psitotzero$ and $\Psitotzero\equiv\Psitot(0)$, where $\Psitot$ is the relative total potential of the main galaxy. For
particles belonging to the satellite galaxy, 
 we assume
\begin{equation}
\alphaIMF(Q)=\CIMF\times(\AIMF\Qtilde+\BIMF),
\label{eq:alphaq2}
\end{equation}
where $\AIMF$ and $\BIMF$ have the same values as for the main galaxy,
$\Qtilde\equiv Q/\Psitotzero$ and $\Psitotzero\equiv\Psitot(0)$, where
now $\Psitot$ is the relative total potential of the satellite
progenitor galaxy. In the special case of equal-mass mergers
($\xistar=1$), we always adopt $\CIMF=1$; for unequal-mass mergers
($\xistar<1$), $\CIMF<1$, i.e.\ the satellite has, on average, lighter
IMF than the main galaxy.  We further assume that during the
simulation each particle maintains its value of $\alphaIMF$ unaltered.
Thus, given the above assumptions, in each binary merger we have three
free dimensionless parameters to assign $\alphaIMF$ to the particles:
$\AIMF$, $\BIMF$ and $\CIMF$. Clearly, the values of $\AIMF$, $\BIMF$
and $\CIMF$ do not influence the dynamical evolution of the
simulation, and can be assumed a posteriori.  It follows that, as far
as the distribution of $\alphaIMF$ is concerned, each of the
considered merging simulations can be formally interpreted in infinite
different ways by choosing the values of $\AIMF$, $\BIMF$ and $\CIMF$
in post-processing.  We stress that the specific forms of
equations~(\ref{eq:alphaq1}) and (\ref{eq:alphaq2}) are not
theoretically justified, but are just simple functions of $Q$ that
allow us to obtain $\alphaIMF$ profiles similar to those measured in
real ETGs (see also \citealt{Cio95} and \citealt{Nip03b}).  Clearly,
$\alphaIMF$ could be also assigned to particles using functions
different from equations~(\ref{eq:alphaq1}) and (\ref{eq:alphaq2}),
for instance higher-degree polynomials of $\Qtilde$, but we found that
the prescriptions~(\ref{eq:alphaq1}) and (\ref{eq:alphaq2}) are
sufficiently general for the purpose of the present investigation (see
\Sect\ref{sec:res}).

\subsection{Diagnostics}
\label{sec:diagn}  

\subsubsection{Spherical systems}

Before considering the analysis of the $N$-body systems of our
simulations (both progenitor galaxies and merger remnants), it is
useful to define a few relevant projected quantities for a spherical
galaxy model with stellar distribution function $f(Q)$.  The stellar
mass surface density profile is
\begin{equation}
\Sigma(R)=\int f(Q)\d \xlos\d^3\vv,
\end{equation}
where $\xlos$ is a coordinate along the line of sight and $R$ is the
projected radius. The effective radius $\Re$ is the projected radius
of a circle containing half of the stellar mass, such that
\begin{equation}
2\pi\int_0^{\Re}\Sigma(R)R\d R=\frac{\Mstar}{2}.
\end{equation}
The \los/ stellar velocity dispersion profile $\sigmalos(R)$ is defined by
\begin{equation}
\sigmalos^2(R)=\frac{1}{\Sigma(R)}\int f(Q)(\vlos-\vbarlos)^2\d \xlos\d^2\vvRv,
\end{equation}
where $\vlos$ is the \los/ velocity, $\vbarlos$ is the mean of $\vlos$
and $\vvRv$ is a vector representing the velocity components in the
plane of the sky. The effective velocity dispersion, i.e.\ the stellar
mass-weighted \los/ central stellar velocity dispersion measured within a
circle of radius $\Re$, is given by
\begin{equation}
\sigmae^2=\frac{2\pi\int_0^{\Re} \Sigma(R)\sigmalos^2(R)R\d R}{\Mstar/2}.
\end{equation}
We note that $\sigmae$ in observed galaxies is a luminosity-weighted
quantity, while here for simplicity we have defined it for our galaxy
models as a mass-weighted quantity. Strictly speaking, the two
definitions differ because we are considering systems with gradients
in $\alphaIMF$ and thus in $\Mstar/L$ \citep[see][]{Ber18}, but this
difference is expected to be small, especially for our models in which
$\sigmalos$ varies at most by $\approx 20\%$ at $R\leq \Re$
(\Sect\ref{sec:sigmalos}).  The projected $\alphaIMF$ profile is
\begin{equation}
\alphaIMF(R)=\frac{1}{\Sigma(R)}\int \alphaIMF(Q)f(Q)\d \xlos\d^3\vv.
\end{equation}
We define the effective IMF mismatch parameter as the mass-weighted projected $\alphaIMF$ within $\Re$, i.e.\
\begin{equation}
\alphae=\frac{2\pi\int_0^{\Re} \Sigma(R)\alphaIMF(R)R\d R}{\Mstar/2}.
\end{equation}

\subsubsection{$N$-body systems}
\label{sec:diagnbody}

The projected properties of our $N$-body models are computed as in
\citet{Nip06} and \citet{NTB09}.  In particular, for any given line of
sight, having determined the ellipticity $\epsilon$ and the principal
axes of the stellar surface density distribution, we measure all the
projected quantities considering concentric elliptical annuli all with
the same ellipticity $\epsilon$. The $k$-th annulus is characterized
by its average circularized radius $R_k$, such that $\log R_{k+1}-\log
R_{k}=\Delta x$ with $\Delta x={\rm constant}$ for all $k$.  The
effective radius $\Re$ is computed as the circularized radius of the
ellipse containing half of the stellar particles in projection.  The
\los/ velocity dispersion at $R_k$ is computed as
\begin{equation}
  \sigmalos(R_k)=\left[\frac{1}{N_k}\sum_i\left(\vlosi-\vbarlosk\right)^2\right]^{1/2},
\end{equation}  
where $\vlosi$ is the \los/ velocity dispersion of the $i$-th
particle, the sum is over all the $N_k$ stellar particles belonging to
the $k$-th projected annulus and $\vbarlosk$ is the mean \los/
velocity of these $N_k$ particles (the total number of stellar
particles is $N=\sum_kN_k$). The effective velocity dispersion is
computed as
\begin{equation}
  \sigmae=\left[\frac{2}{N}\sum_i\left(\vlosi-\vbarlos\right)^2\right]^{1/2},
\end{equation}
where $\vbarlos$ is the mean \los/ velocity of the $N/2$ stellar
particles contained within the ellipse with circularized radius $\Re$
and the sum is over the same $N/2$ particles.

We compute the radial profile of the IMF mismatch parameter as
\begin{equation}
\alphaIMF(R_k)=\frac{1}{N_k}\sum_i\alphaIMFi,
\end{equation}  
where $\alphaIMFi$ is the value of $\alphaIMF$ of the $i$-th particle,
and the sum is over all the $N_k$ stellar particles belonging
to the $k$-th annulus. The effective IMF mismatch parameter is
computed as
\begin{equation}
\alphae=\frac{2}{N}\sum_i\alphaIMFi,
\end{equation}  
where the sum is over all the $N/2$ stellar particles
contained within the ellipse with circularized radius $\Re$.

For each $N$-body system we consider 50 projections with different
(random) lines of sight and we compute the mean and standard deviation
of $\sigmalos(R_k/\Re)$, $\sigmae$, $\alphaIMF(R_k/\Re)$ and
$\alphae$.

\subsection{Physical units and normalizations}
\label{sec:units}

Given the scale-free nature of gravity, our dissipationless merging
simulations are fully scalable in mass and length.  We take as mass
unit $\Mu=\Mstarmain$ the stellar mass of the main galaxy and as
length unit $\lu=\rstarmain$ the stellar scale radius of the main
galaxy. Thus, for the adopted main galaxy model
(\Sect\ref{sec:progenitor}), the effective radius of the main galaxy
is
\begin{equation}
  \Remain\simeq 1.23\left(\frac{\lu}{\kpc}\right)\kpc
\end{equation}  
and the effective velocity dispersion of the main galaxy is 
\begin{equation}
  \sigmaemain\simeq 92.9\left(\frac{\Mu}{10^{10}\Msun}\right)^{1/2}\left(\frac{\lu}{\kpc}\right)^{-1/2}\kms.
\end{equation}  
As far as the IMF mismatch parameter is concerned, we normalize all
our results to $\alphaemain$, the value of $\alphae$ of the main
galaxies. The value of $\alphaemain$ is independent of both $\Mu$ and
$\lu$, and can be chosen freely if one wants to apply our results to
specific observational targets. We recall that for an ensemble of
stellar particles, the quantity $\alphaIMF$ used in this paper is the
true $\Mstar/L$ normalized to the $\Mstar/L$ of a reference IMF
(\Eq\ref{eq:alphaimf}).  Our results can be interpreted by choosing
freely the reference IMF: for instance a Salpeter, a Kroupa or a
Chabrier IMF.

\section{Results}
\label{sec:res}

\begin{figure} 
  \centerline{\psfig{file=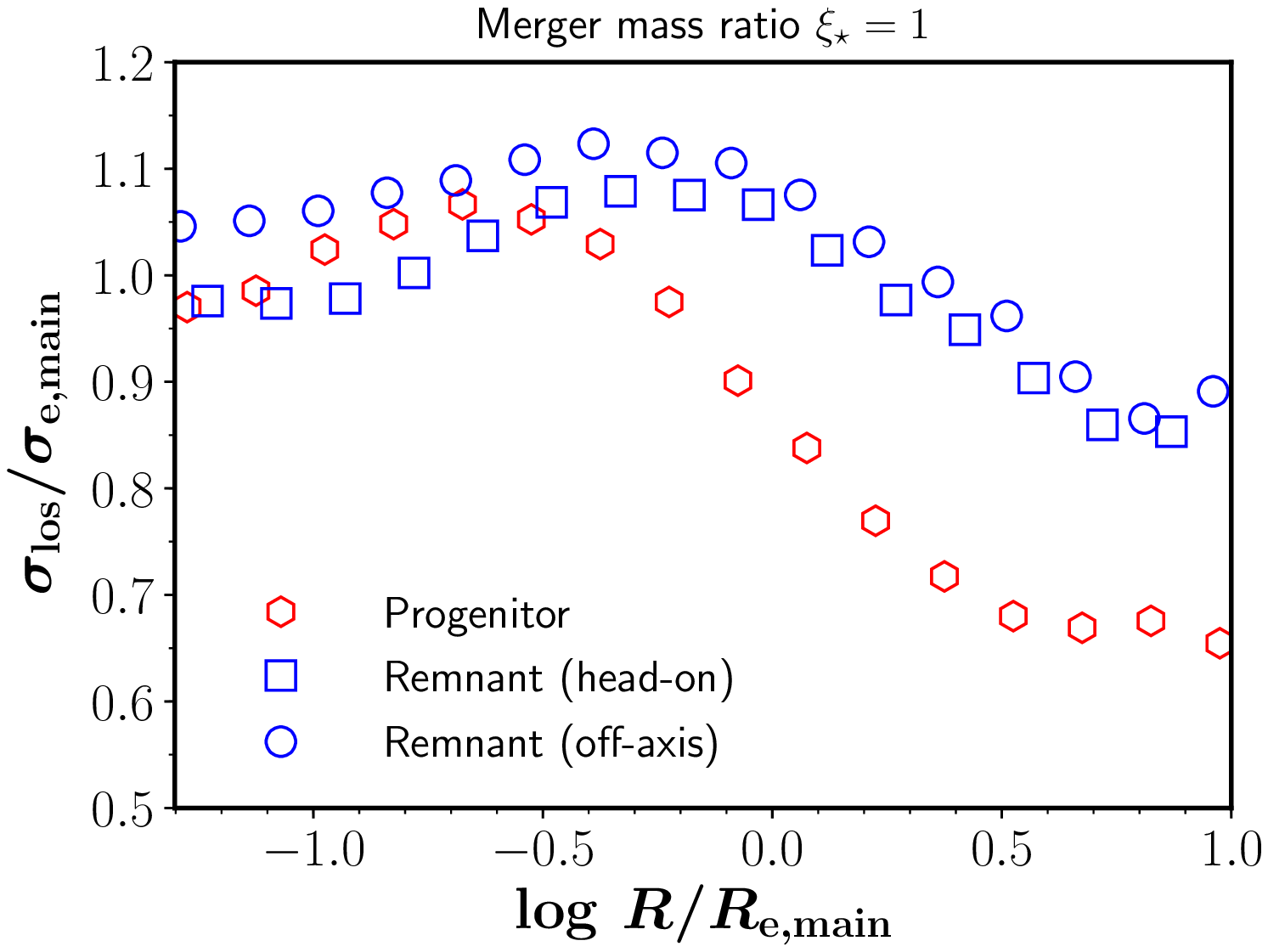,width=\hsize}}
  \centerline{\psfig{file=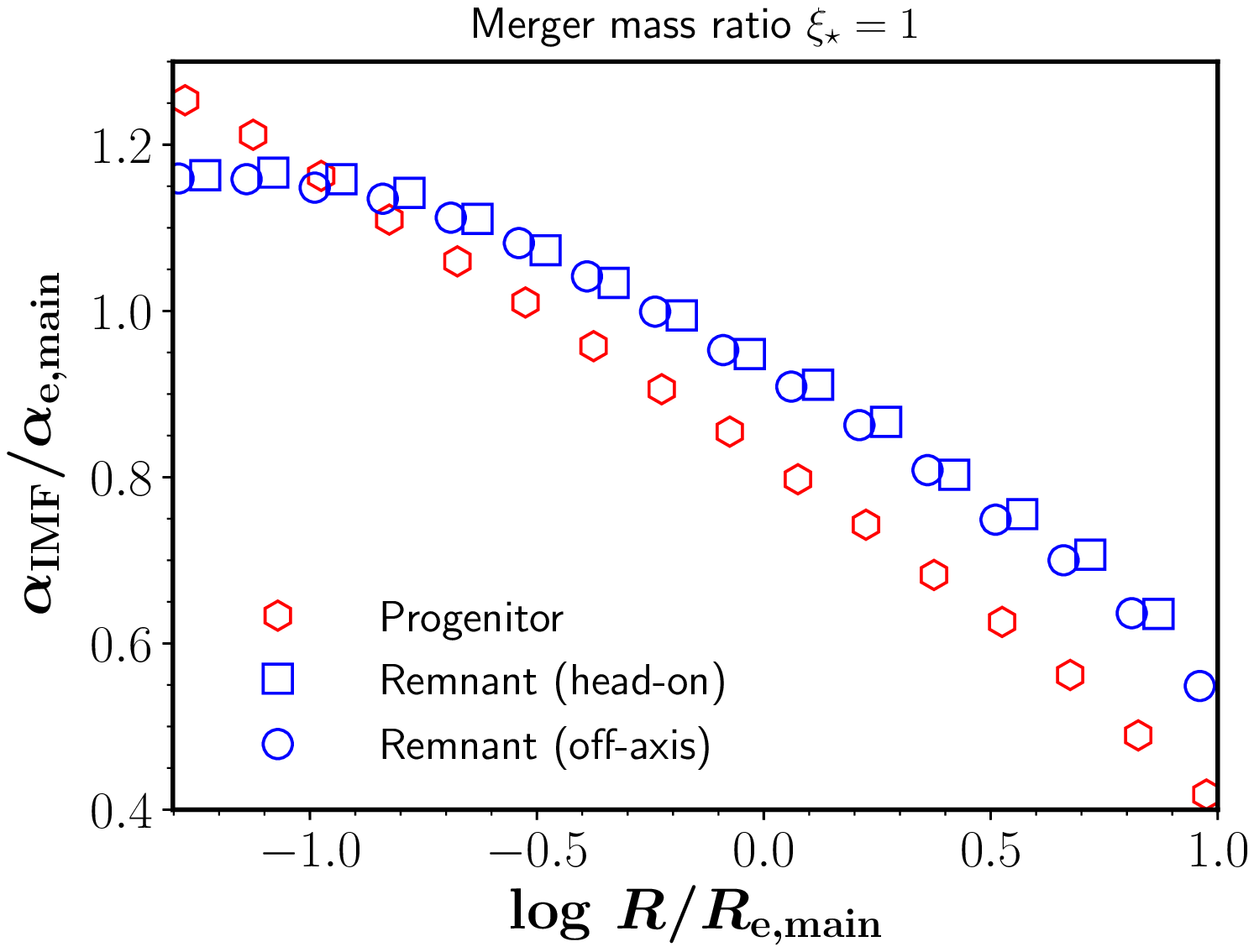,width=\hsize}}
  \caption{Angle-averaged \los/ velocity dispersion (upper panel) and
    IMF mismatch parameter (lower panel) profiles of the remnants of
    the equal-mass merging simulations 1Dh (squares) and 1Do
    (circles), and of their progenitor galaxies (hexagons).
    $\Remain$, $\sigmaemain$ and $\alphaemain$ are, respectively, the
    effective radius, velocity dispersion and IMF mismatch parameter
    of the progenitor galaxies.}
\label{fig:rimf_xi1}
\end{figure}

\begin{figure} 
  \centerline{\psfig{file=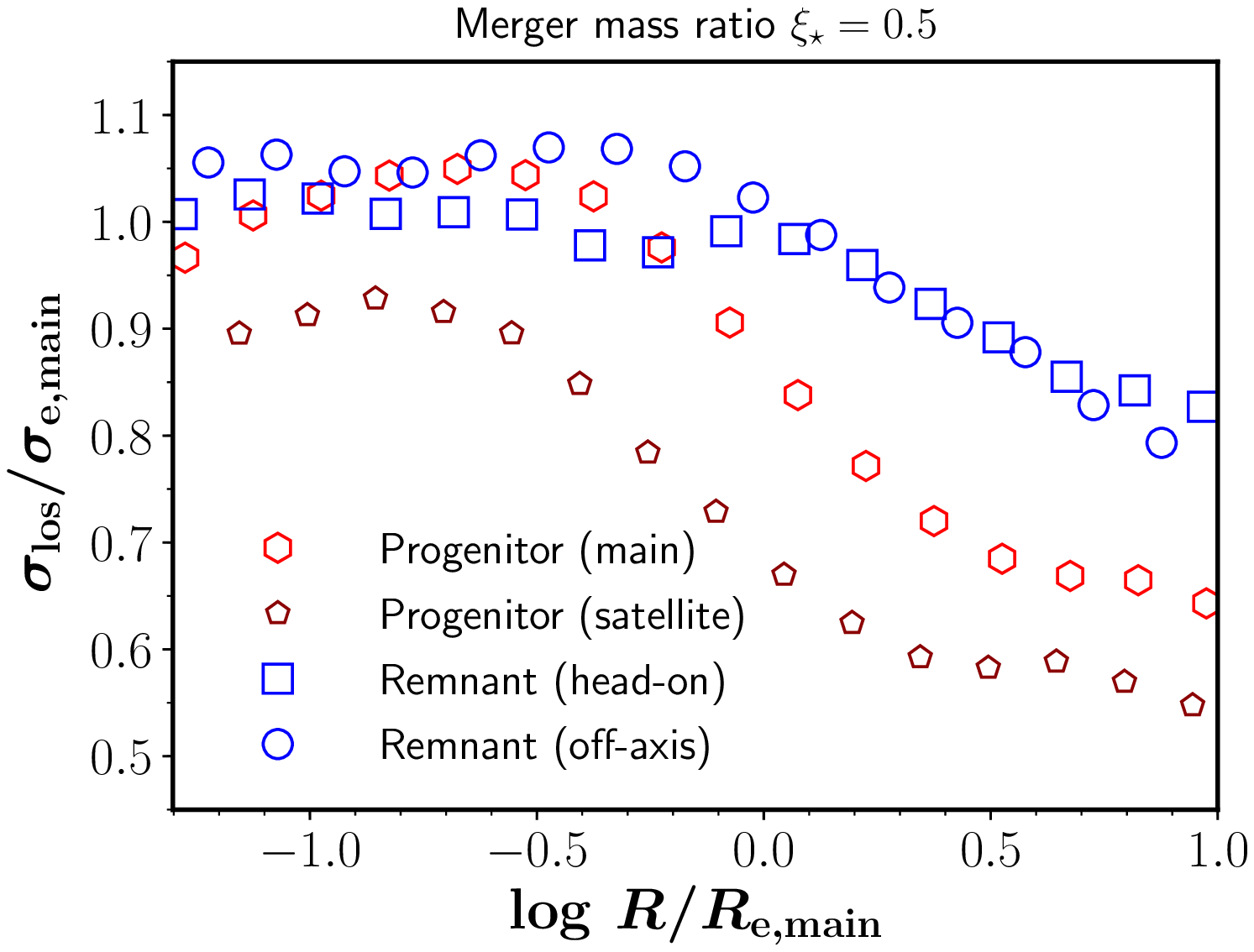,width=\hsize}}
  \centerline{\psfig{file=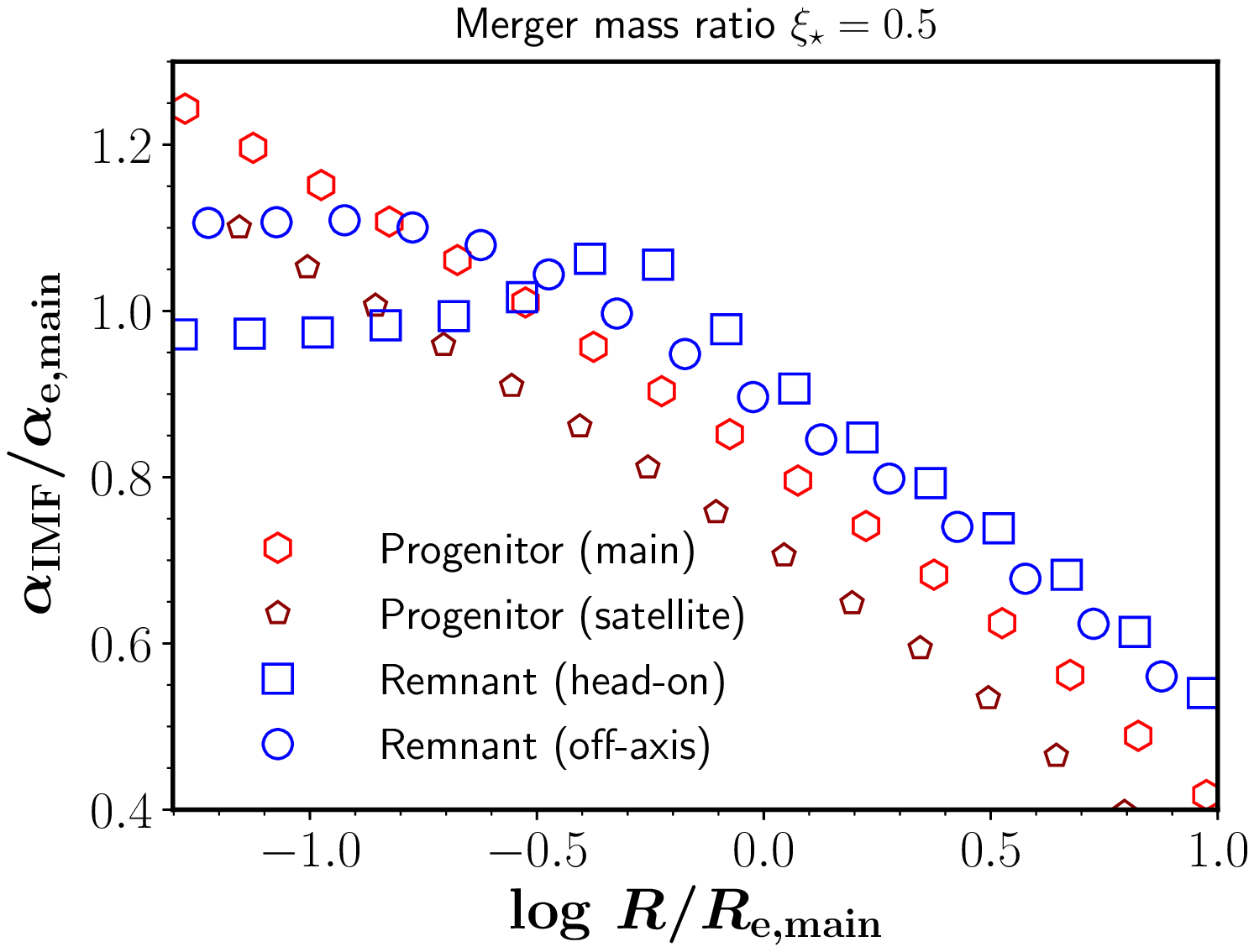,width=\hsize}}
  \caption{Same as Fig.\ \ref{fig:rimf_xi1}, but for the merging
    simulations with stellar mass ratio $\xistar=0.5$. The hexagons
    and pentagons represent, respectively, the main and satellite
    progenitor galaxies. The remnants are indicated with squares
    (simulation 0.5Dh) and circles (simulation 0.5Dh).  $\Remain$,
    $\sigmaemain$ and $\alphaemain$ are, respectively, the effective
    radius, velocity dispersion and IMF mismatch parameter of the main
    progenitor galaxy. }
  \label{fig:rimf_xi05}
\end{figure}

\begin{figure} 
  \centerline{\psfig{file=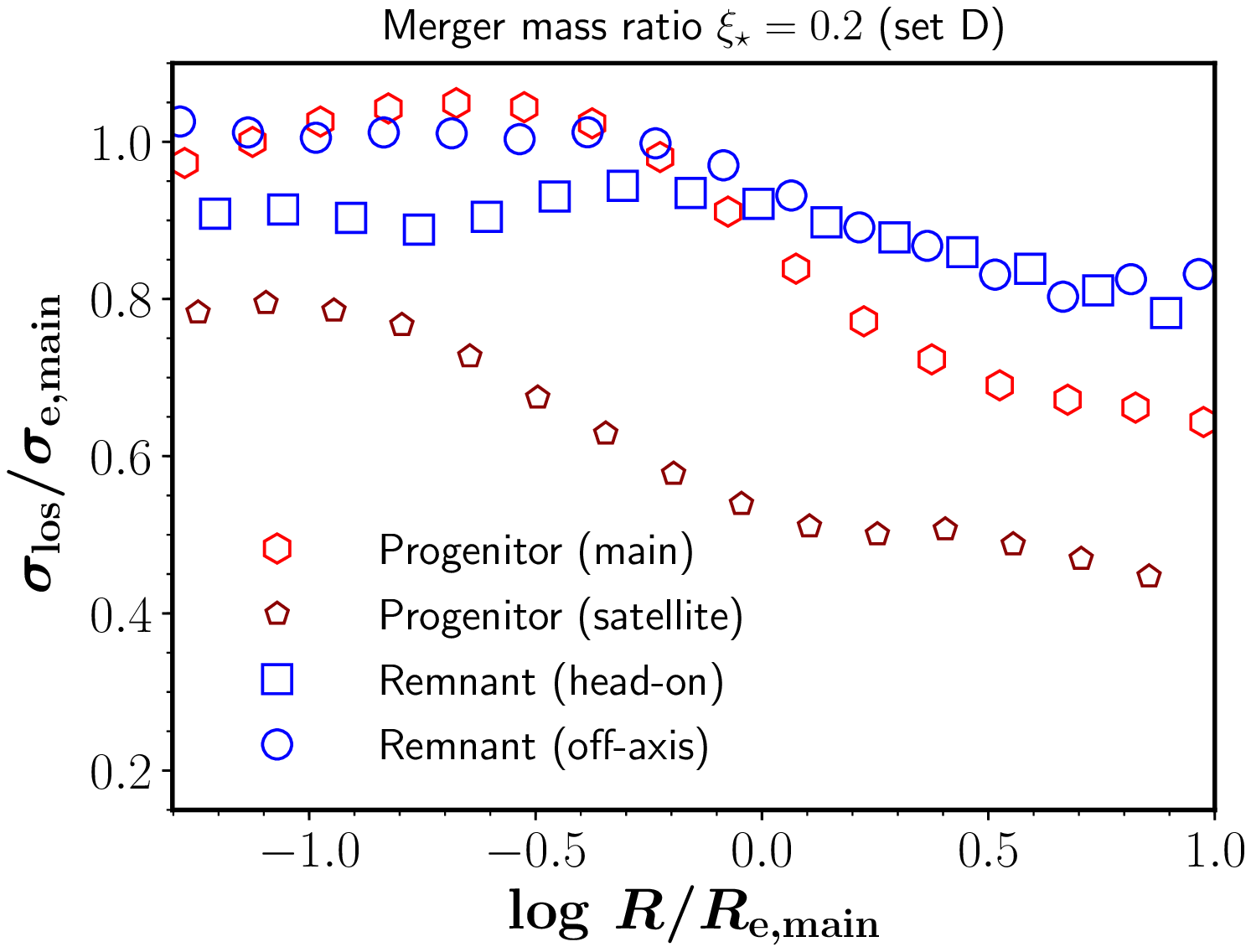,width=\hsize}}
  \centerline{\psfig{file=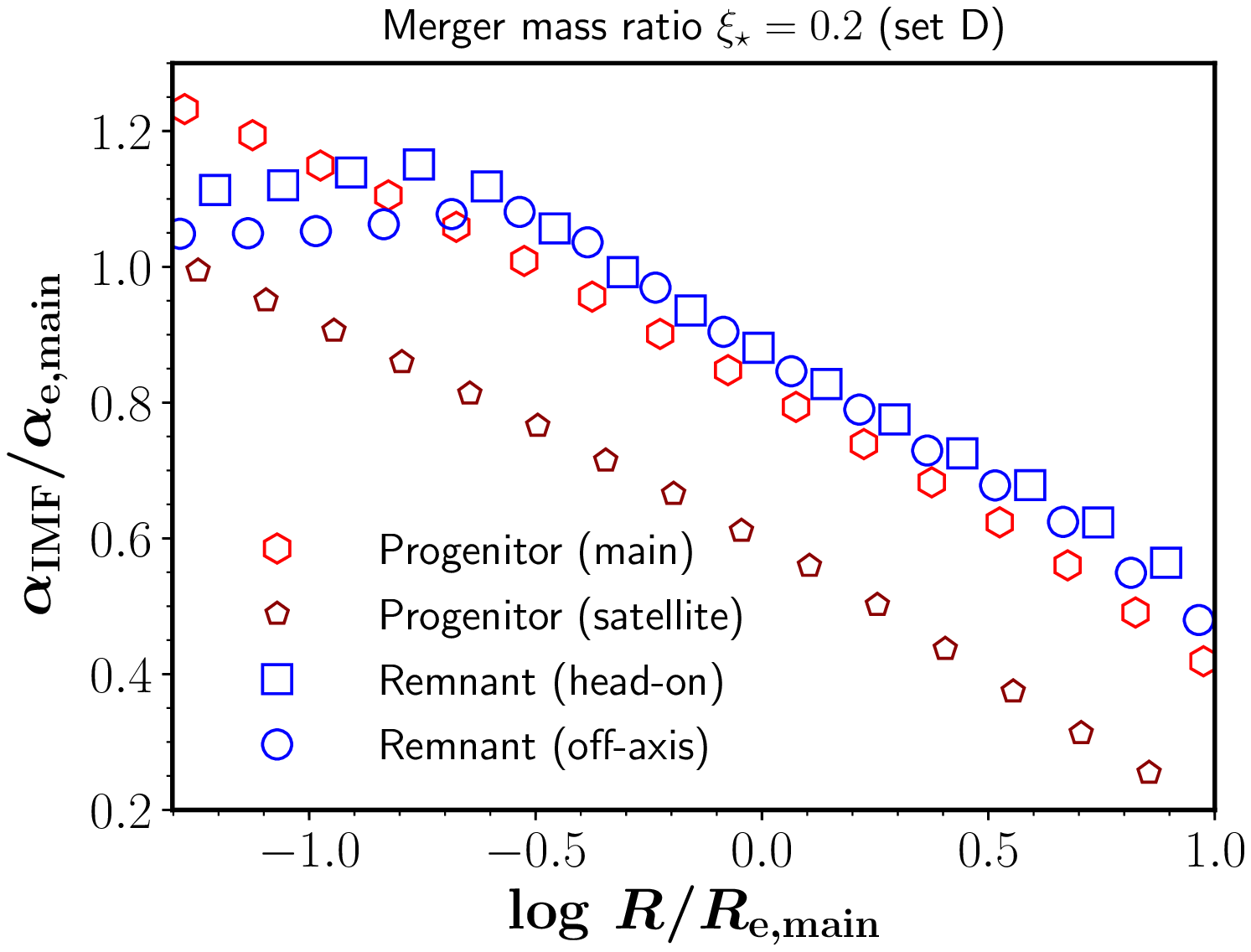,width=\hsize}}
\caption{Same as Fig.\ \ref{fig:rimf_xi05}, but for the remnants of
  the $\xistar=0.2$ merging simulations with stellar mass ratio 0.2D3h
  (squares) and 0.2D3o (circles), and for their main (hexagons) and
  satellite (pentagons) progenitor galaxies.}
\label{fig:rimf_xi02}
\end{figure}

\begin{figure} 
  \centerline{\psfig{file=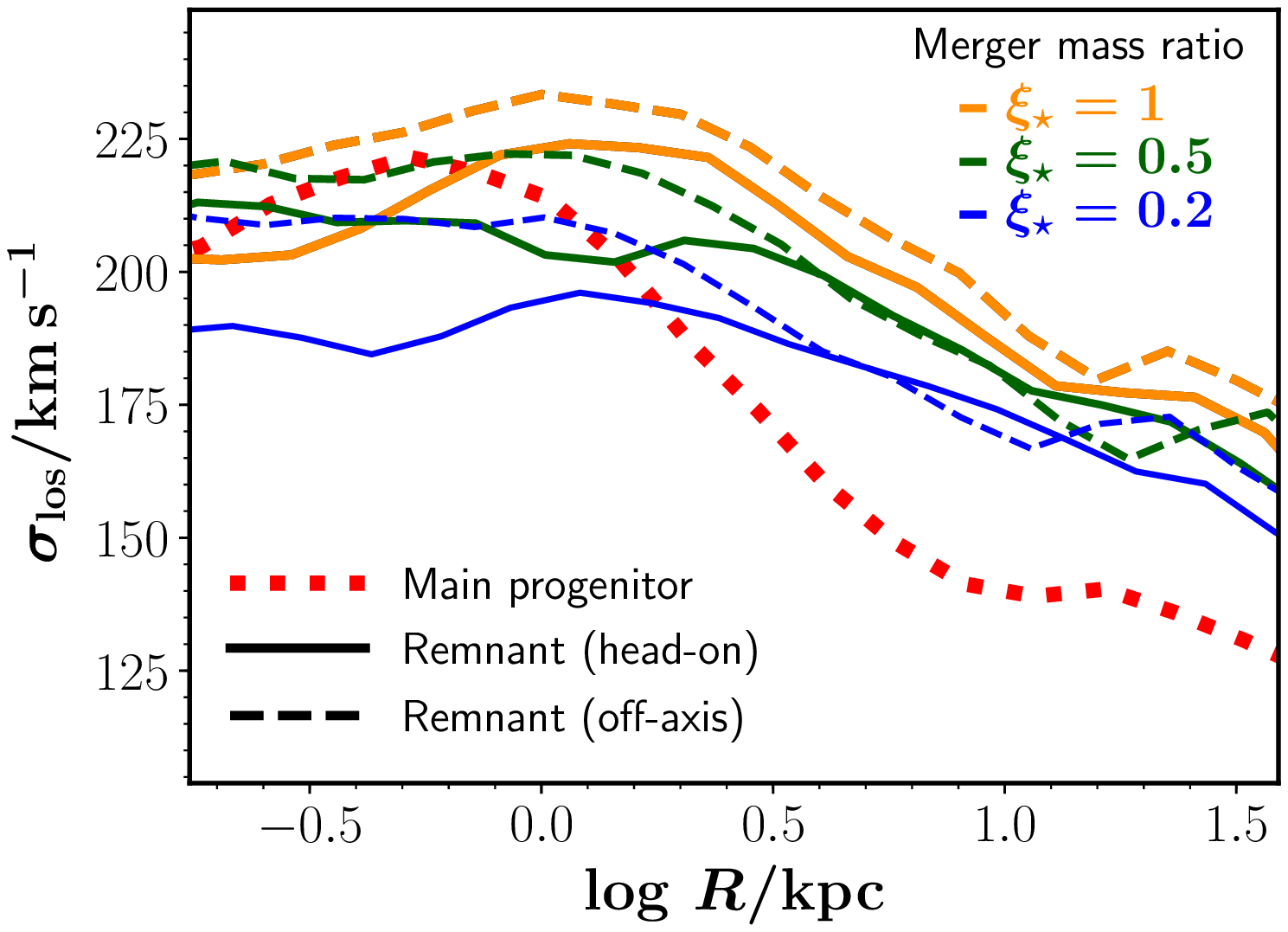,width=\hsize}}
  \centerline{\psfig{file=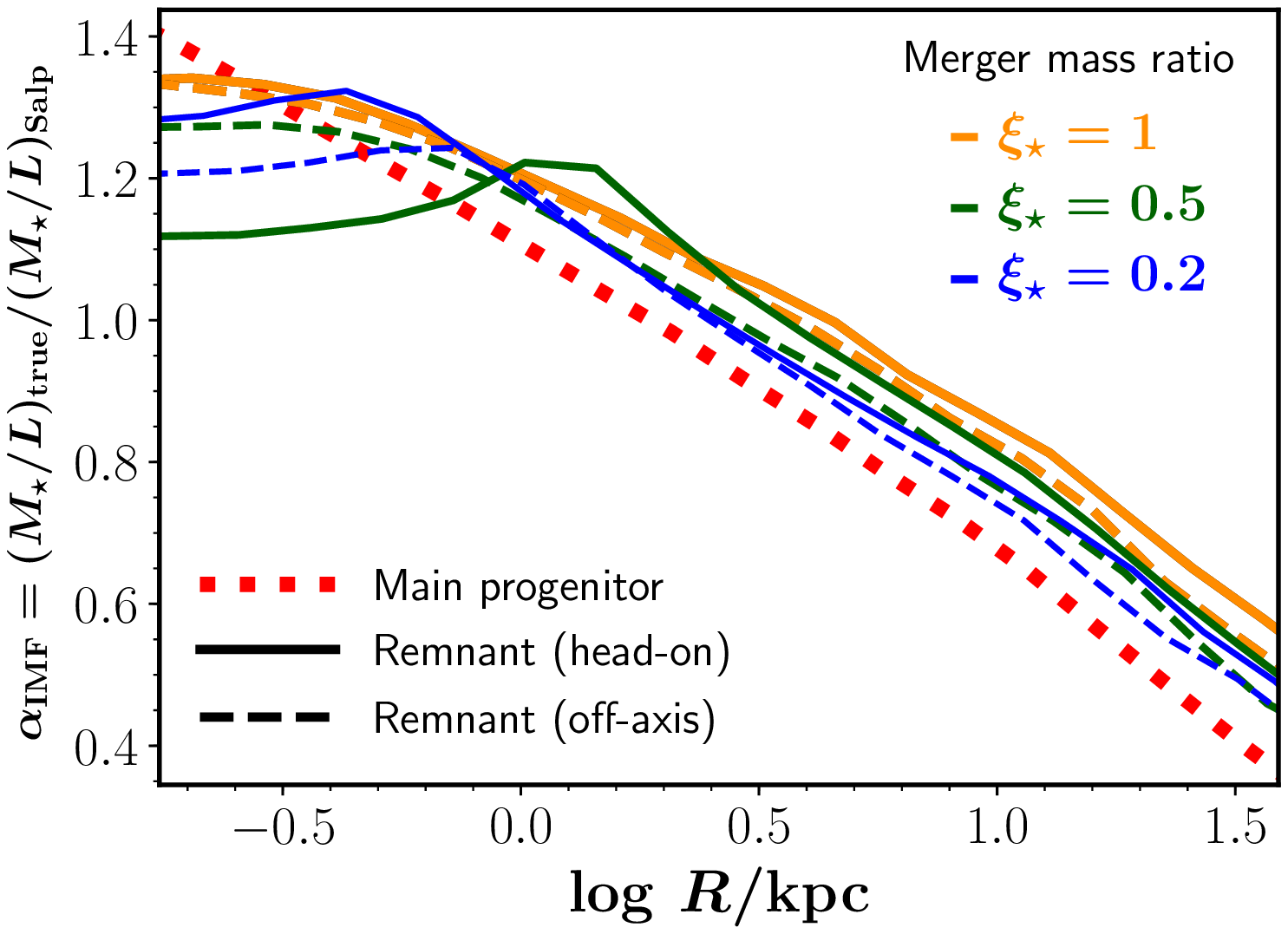,width=\hsize}}
  \caption{Angle-averaged \los/ velocity dispersion (upper panel) and
    IMF mismatch parameter (lower panel) profiles of the remnants of
    the head-on (solid curves) and off-axis (dashed curves) merging
    simulations, and of their main progenitor galaxy (dotted curve) in
    physical units, assuming mass unit $\Mu=10^{11}\Msun$, length unit
    $\lu=2\kpc$ (so $\Mstarmain=10^{11}\Msun$, $\Remain\simeq2.5\kpc$
    and $\sigmaemain\simeq 208\kms$; see \Sect\ref{sec:units}) and
    $\alphaemain=1.15$, taking as reference the Salpeter IMF. The
    curves representing the remnants are thicker for higher merger
    mass ratio ($\xistar=1$, orange curves; $\xistar=0.5$, green
    curves; $\xistar=0.2$, blue curves).}
\label{fig:rimf}
\end{figure}

\begin{figure} 
  \centerline{\psfig{file=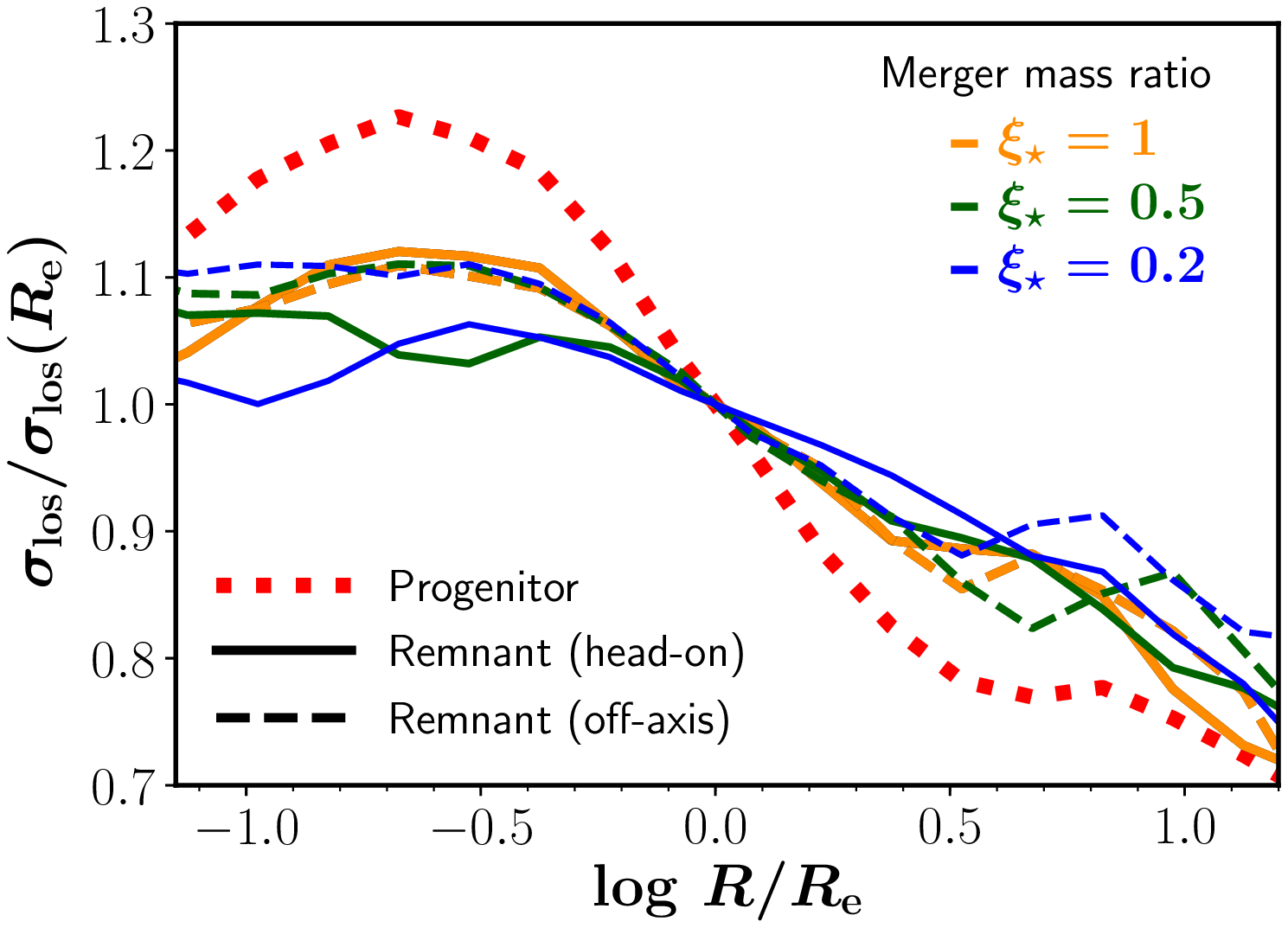,width=\hsize}}
  \centerline{\psfig{file=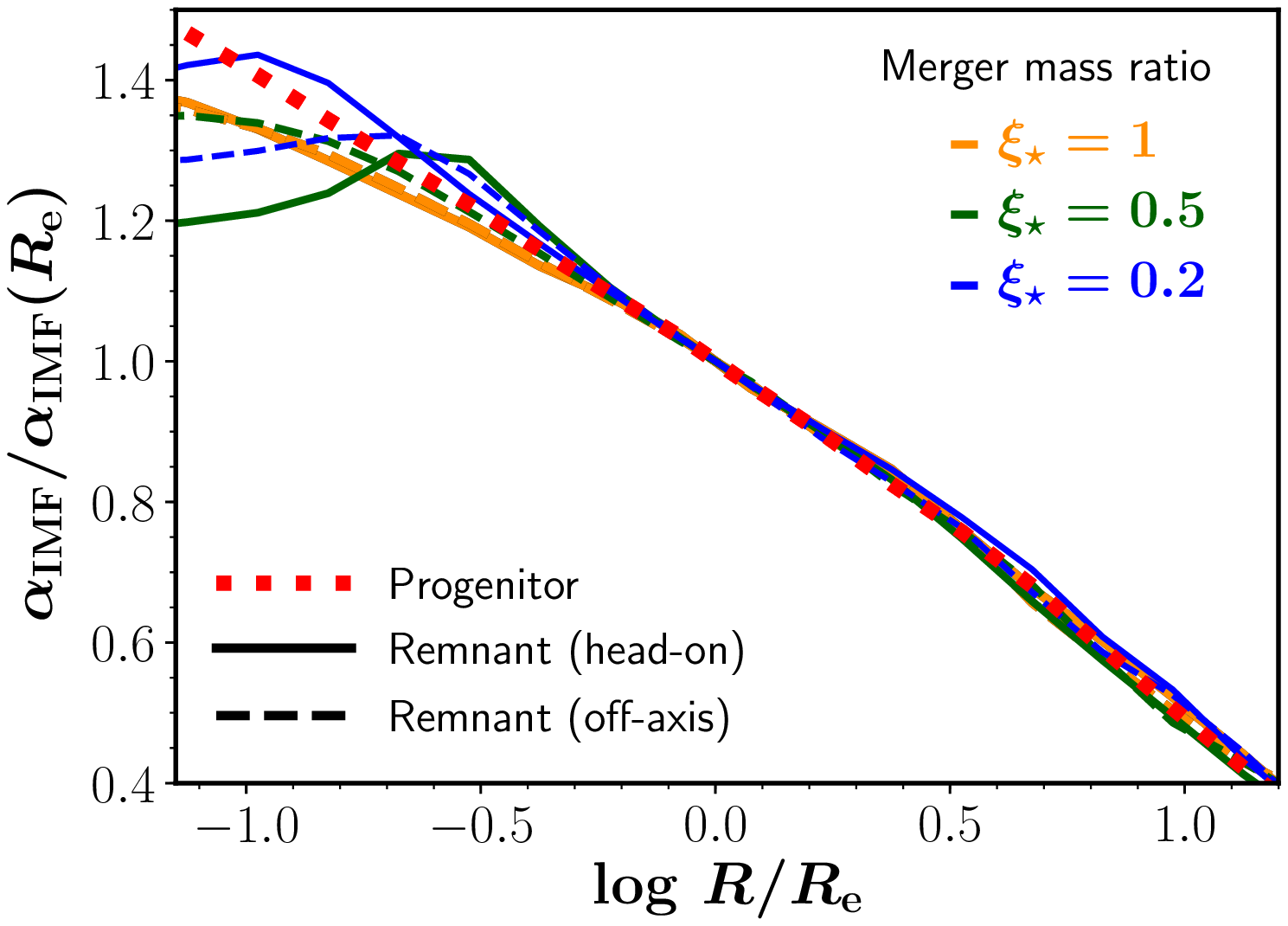,width=\hsize}}
  \caption{Same as \Fig\ref{fig:rimf}, but here the profiles are
    normalized to the values of $\sigmalos$ and $\alphaIMF$ at the
    effective radius $\Re$ ($\Re=\Remain$ for the main progenitor and
    $\Re=\Reremn$ for the remnants).}
\label{fig:rimf_norm}
\end{figure}

Here we present the results on the evolution of $\alphaIMF$ in mergers
based on the dissipationless binary merging simulations described in
\Sect\ref{sec:simu}. In order to assign values of $\alphaIMF$ to
particles, we have to choose, in post-processing, the values of the
coefficients $\AIMF$, $\BIMF$ and $\CIMF$
(\Sect\ref{sec:assignalpha}). We present here results obtained
adopting $\AIMF=2$, $\BIMF=0$ and $\CIMF$ that depends on $\xistar$
(the specific values of $\CIMF$ are reported in
\Tab\ref{tab:simu}). In \Sects\ref{sec:imfprofiles} and
\ref{sec:localphasigma} we show that with this choice we obtain
progenitor galaxy models with realistic $\alphaIMF$ profiles.  We
experimented with different choices of the values of $\AIMF$, $\BIMF$
and $\CIMF$, finding that, provided that the values give realistic
$\alphaIMF$ profiles in the progenitors, the merger-driven evolution
of the $\alphaIMF$ profiles is weakly dependent on the specific choice
of $\AIMF$, $\BIMF$ and $\CIMF$. In Appendix~\ref{sec:varimf}, we show
examples illustrating the effect of choosing values of $\AIMF$ and
$\BIMF$ giving significantly different $\alphaIMF$ profiles for the
progenitors.  The results presented in this section refer to
simulations (set D in \Tab\ref{tab:simu}) in which the main and
satellite galaxies are structurally and dynamically homologous (see
\Sect\ref{sec:set}).  In Appendix~\ref{sec:varstruct} we compare some
of these simulations with analogous simulations in which the satellite
and the progenitor galaxies are not homologous, finding results very
similar to those presented in this section.

The profiles of $\sigmalos$ and $\alphaIMF$, as well as the integrated
quantities $\sigmae$ and $\alphae$, depend on the line of sight. In
this section, we show for each $N$-body system a single value of each
of these quantities, that is the mean over 50 different lines of
sight. In some cases we associate to this mean, as an error bar, the
corresponding standard deviation (see \Sect\ref{sec:diagnbody}).

\subsection{Velocity dispersion profiles}
\label{sec:sigmalos}

The stellar \los/ velocity dispersion profiles of the progenitor
galaxies and of the remnants of the simulations are shown in the upper
panels of \Figs\ref{fig:rimf_xi1}-\ref{fig:rimf_xi02}.  In the main
galaxy (hexagons in the plots) $\sigmalos$ increases from the centre
out to $\approx 0.25\Re$, where it peaks, and decreases outwards at
larger radii out to $\approx 10\Re$. The positive inner gradient of
$\sigmalos$ is usually not observed in real ETGs
\citep[e.g.][]{deZ02,Fal17}, but we note that in our model such a
positive gradient is limited to very small radii ($R\lesssim0.2\Re$),
not always probed in spectroscopic observations. In fact, the inner
$\sigmalos$ profile of the progenitors is very shallow ($\sigmalos$
increases by $\approx 10\%$ from $R\approx 0.02\Re$ to $R\approx
0.2\Re$), so we do not expect this feature of our progenitor galaxy
models to affect the main conclusions of our work. From the peak to
$\Re$, the decrease in $\sigmalos$ is about 20\%, broadly consistent
with the observed profiles of ETGs \citep[see e.g.][]{Ras14,Vea18}.
The satellite galaxies in the unequal-mass mergers (pentagons in
\Figs\ref{fig:rimf_xi05}-\ref{fig:rimf_xi02}) are homologous to the
main galaxy, so they have $\sigmalos$ profiles with the same shape as
that of the main, but just scaled down to lower values of velocity
dispersion.

The remnants of the equal-mass mergers (\Fig\ref{fig:rimf_xi1}) have
\los/ velocity dispersion comparable to or higher than that of the
progenitors at all radii: we find that $\sigmalos$ increases more in
the off-axis than in the head-on merger, as found in general in
dissipationless mergers \citep[e.g.][]{Boy06,NTB09}. The remnants of
unequal-mass mergers (\Figs\ref{fig:rimf_xi05}-\ref{fig:rimf_xi02})
have lower $\sigmalos$ profiles than those of the equal-mass mergers.
The lowest $\sigmalos$ profile is obtained for the remnant of the
head-on minor merger ($\xistar=0.2$). Focusing on the central parts of
the velocity dispersion profiles, we notice that the behaviour of
$\sigmalos$ in the remnant ranges from being higher (as in the
off-axis $\xistar=1$ merger) to being lower (as in the head-on
$\xistar=0.2$ merger) than $\sigmalos$ in the main progenitor.

In \Figs\ref{fig:rimf_xi1}-\ref{fig:rimf_xi02} $\sigmalos$ and $R$ are
normalized, respectively, to the effective velocity dispersion
$\sigmaemain$ and to the effective radius $\Remain$ of the main
progenitor galaxy for both the progenitors and the remnants, so all
the $\sigmalos$ profiles are shown on the same physical scale. These
profiles can be converted in physical units by fixing $\Mu$ and $\lu$
(see \Sect\ref{sec:units}): an example is given in the upper panel of
\Fig\ref{fig:rimf}, showing the $\sigmalos$ profiles of the main
progenitor and of the remnants of the same six simulations as in
\Figs\ref{fig:rimf_xi1}-\ref{fig:rimf_xi02}.  When interpreting the
variations in the $\sigmalos$ profiles shown in
\Figs\ref{fig:rimf_xi1}-\ref{fig:rimf} one must bear in mind that, as
well known, dry mergers make galaxies more diffuse (see values of
$\Reremntilde$ in \Tab\ref{tab:simu}), so the effect of mergers is not
only to change the shape of the $\sigmalos$ profiles, but also to
``stretch'' them horizontally. To isolate the effect of the mergers on
the shape of the $\sigmalos$ profiles, it is useful to normalize $R$
and $\sigmalos$ of each system (either progenitor or remnant) to,
respectively, their own $\Re$ and $\sigmalos$ at $\Re$, as done in the
upper panel of \Fig\ref{fig:rimf_norm}. The main effect of the mergers
on the shape of the $\sigmalos$ profile is an overall flattening of
the profile. We also note that the positive central $\sigmalos$
gradient of the progenitor tends to be erased by the merging.

\subsection{Initial mass function mismatch parameter profiles}
\label{sec:imfprofiles}

The stellar IMF mismatch parameter profiles of the progenitor galaxies
and of the remnants of the simulations are shown in the lower panels
of \Figs\ref{fig:rimf_xi1}-\ref{fig:rimf_xi02}.  In the progenitor
galaxies (hexagons and pentagons in the plots) $\alphaIMF$ decreases
monotonically with radius, by a factor of about two from the centre to
$\approx 3\Re$, consistent with the observational estimates of
present-day ETGs (e.g.\ \citealt{LaB19}). We note that, given that by
construction $\AIMF$ and $\BIMF$ are the same in the two progenitor
galaxies, the $\alphaIMF$ profile of the satellite galaxy (pentagons)
is just a scaled-down version of that of the main galaxy (hexagons):
specifically, at given $R/\Re$, $\alphaIMFsat=\CIMF\alphaIMFmain$,
with $\CIMF\leq 1$.

Let us focus first on equal-mass mergers (\Fig\ref{fig:rimf_xi1}). The
merger remnant has, with respect to the progenitor galaxies, a
shallower central gradient of $\alphaIMF$. The remnant's $\alphaIMF$
profile is very similar for off-axis (circles) and head-on (squares)
encounters. Moving to unequal-mass major ($\xistar=0.5$;
\Fig\ref{fig:rimf_xi05}) and minor ($\xistar=0.2$;
\Fig\ref{fig:rimf_xi02}) mergers, we note that the main effect of the
merger is to flatten the central gradient of $\alphaIMF$ by producing
a central `core' of constant $\alphaIMF$ with size ranging from
$\approx0.1$ to $\approx0.3$ in units of $\Remain$. The core is
produced by the lower-$\alphaIMF$ stellar particles of the satellite
that settle in the central regions and mix with the higher-$\alphaIMF$
stellar particles of the main. For major mergers ($\xistar=0.5$) the
core is larger for head-on (squares) than for off-axis (circles)
encounters, while for minor mergers ($\xistar=0.2$) the core is larger
for off-axis than for head-on encounters, which suggests that mixing
and accretion of satellite's stars in the central region of the
remnant depends in a non-trivial way both on the orbit of the
encounter and on the merger mass ratio, for given structural
properties of the progenitors.

The $\alphaIMF$ profiles shown in
\Figs\ref{fig:rimf_xi1}-\ref{fig:rimf_xi02} are all on the same
physical scale, because $\alphaIMF$ and $R$ are normalized,
respectively, to the effective IMF mismatch parameter $\alphaemain$
and to effective radius $\Remain$ of the main progenitor.  In the
lower panel of \Fig\ref{fig:rimf}, the $\alphaIMF$ profiles of the
main progenitor and of the remnants are shown in physical units, for a
representative case in which we have assumed $\alphaemain=1.15$ and
the Salpeter IMF as reference. As done for the $\sigmalos$ profiles
(\Sect\ref{sec:sigmalos}), in \Fig\ref{fig:rimf_norm} (lower panels)
we show the same $\alphaIMF$ profiles as in \Fig\ref{fig:rimf}, but
normalized to $\Re$ and $\sigmalos(\Re)$, to highlight the variations
in the shape of the profiles, which are significant only at
$R\lesssim\Re$.  In \Figs\ref{fig:rimf_xi1}-\ref{fig:rimf} the
remnants have higher $\alphaIMF$ than the progenitor at large radii
mainly because of the merger-driven size evolution: on average, stars
originally belonging to the main progenitor orbit at larger (physical)
radius in the remnant.

Overall, under the considered hypotheses (heavier average IMF in more
massive progenitors and negative radial gradients of $\alphaIMF$ in
the progenitors), the effect of dissipationless mergers is invariably
to redistribute $\alphaIMF$ by reducing it in the central regions and
slightly increasing it at larger radii, thus weakening the $\alphaIMF$
gradient. We note that this weakening of the $\alphaIMF$ gradient is
not a necessary consequence of merging with a lower-$\alphaIMF$
satellite, but it must be expected when the $\alphaIMF$ gradient in
the main progenitor is sufficiently strong and when the satellite is
sufficiently compact not to be completely disrupted in the outskirts
of the main.  If the main progenitor had a negligible $\alphaIMF$
gradient, a merger with a low $\alphaIMF$ satellite could give rise to
$\alphaIMF$ profiles with both positive and negative $\alphaIMF$
gradients (see \Sect\ref{sec:intro}).  The effect of dry merging on
the $\alphaIMF$ profiles depends also on the structural properties of
the satellite (see \Sect\ref{sec:hier}). The accretion of loose
(low-density) satellites with low $\alphaIMF$ can produce a negative
radial gradient of $\alphaIMF$, because such satellites tend to
deposit their stars mainly in the outer regions of the remnant, while
the central $\alphaIMF$ is determined by stars formed in situ
\citep[see][for a discussion]{Smi20}.

\begin{figure}
  \centerline{\psfig{file=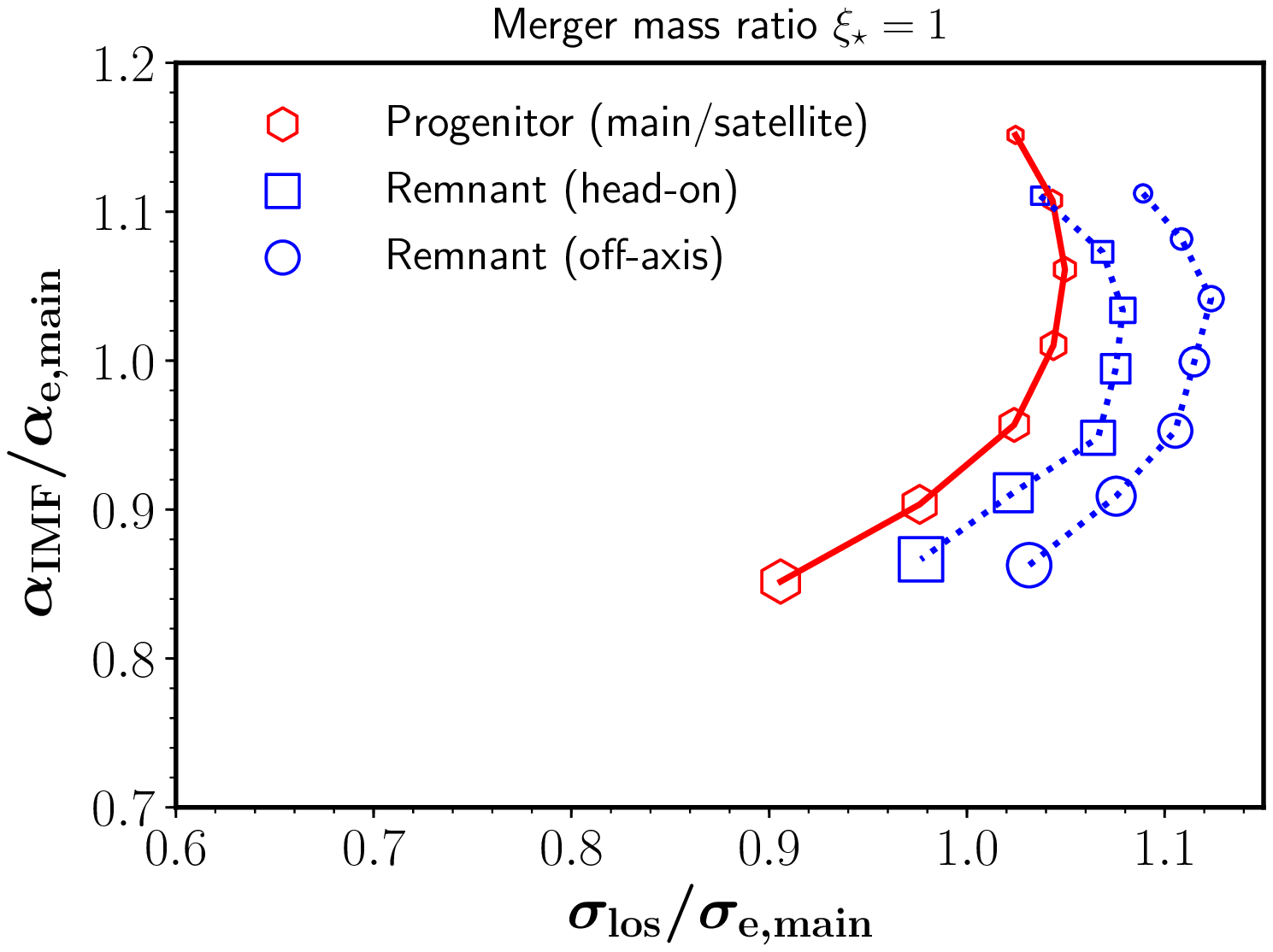,width=\hsize}}  
  \centerline{\psfig{file=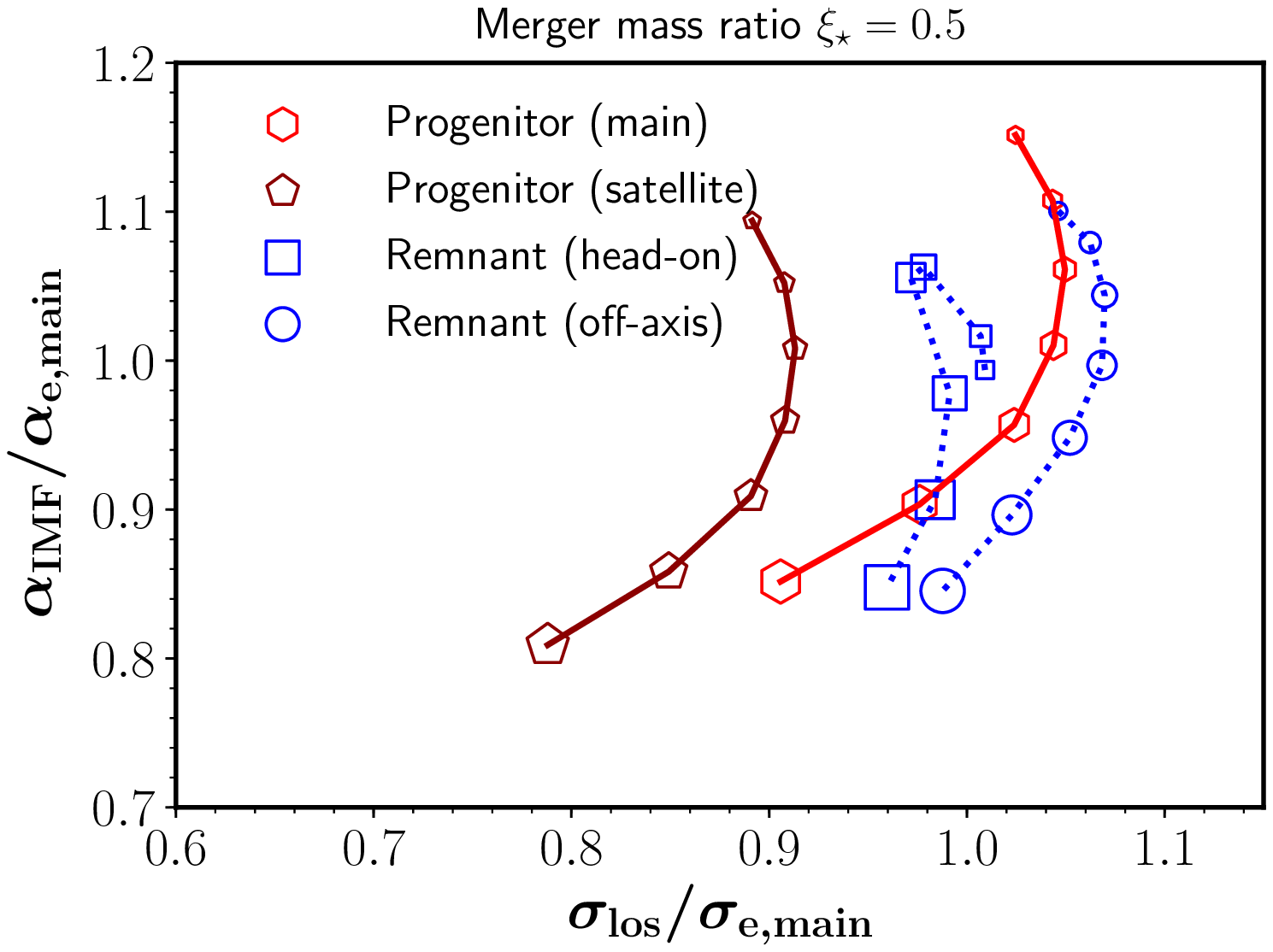,width=\hsize}}  
  \centerline{\psfig{file=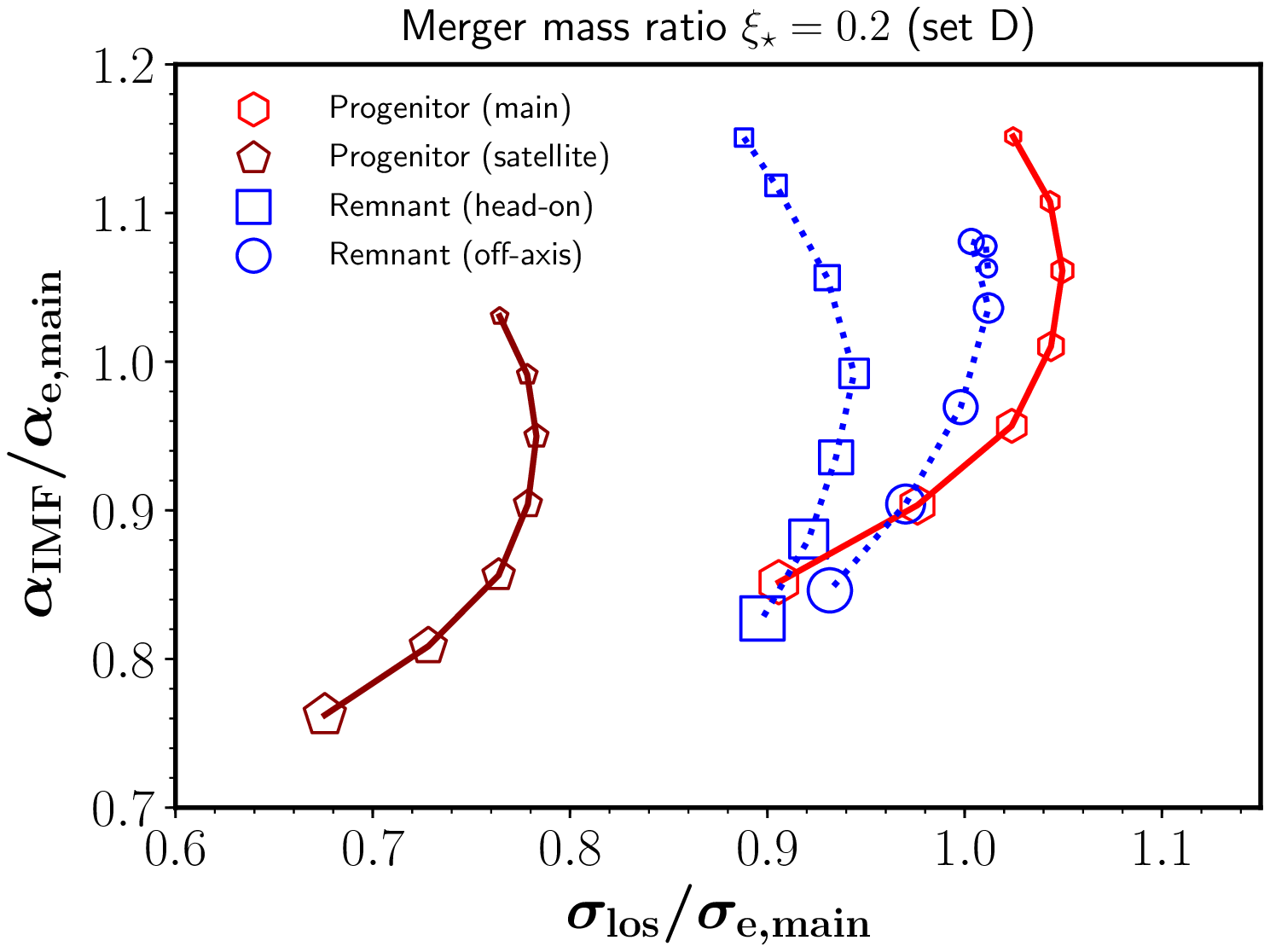,width=\hsize}}
  \caption{Local (i.e.\ at given $R$) IMF mismatch parameter
    $\alphaIMF$ as a function of local (at the same $R$) \los/
    velocity dispersion $\sigmalos$ of the progenitor galaxies and of
    the remnants of simulations 1Dh and 1Do ($\xistar=1$, top panel),
    0.5Dh and 0.5Do ($\xistar=0.5$, middle panel) and 0.2Dh and 0.2Do
    ($\xistar=0.2$, bottom panel) in the radial interval $0.1\leq
    R/\Re\leq1$, where $\Re=\Remain$ for the main galaxy, $\Re=\Resat$
    for the satellite galaxy and $\Re=\Reremn$ for the
    remnants. Symbols and normalizations are the same as in
    Figs.\ \ref{fig:rimf_xi1}-\ref{fig:rimf_xi02}. The size of the
    symbol increases for increasing $R$.}
\label{fig:sigimf}
\end{figure}

\subsection{Local $\alphaIMF$ as a function of local $\sigmalos$}
\label{sec:localphasigma}

In their study of the spatially resolved stellar IMF, \citet{Par18}
explored the distribution of the local $\alphaIMF$ as a function of
the local $\sigmalos$ for a large sample of ETGs (see figure 16 in
that paper; see also \citealt{Dom19,Dom20}). Here we perform a similar
analysis, but for our merging simulations. In \Fig\ref{fig:sigimf} we
plot the local $\alphaIMF$ as a function of the local $\sigmalos$ for
the remnants and the progenitor galaxies of our simulations with
$\xistar=1$ (top panel), $\xistar=0.5$ (middle panel) and
$\xistar=0.2$ (bottom panel).  Similar to \citet{Par18}, in these
diagrams we plot only values of $\sigmalos(R)$ and $\alphaIMF(R)$
lying in the radial range $0.1\lesssim R/\Re\lesssim 1$. The bottom
parts of the distributions correspond to $R\approx \Re$ (lower values
of $\alphaIMF$), while the top parts to the central regions (higher
$\alphaIMF$). The $\alphaIMF$-$\sigmalos$ distribution of the
progenitor galaxies of our simulations is qualitatively similar to
that inferred for real ETGs by \citet[][]{Par18}, though,
quantitatively, the relative variations in $\sigmalos$ and $\alphaIMF$
are somewhat smaller in our model galaxies than in the ETGs of Parikh
et al.'s sample. In both model and real ETGs the trend is that
$\alphaIMF$ tends to increase with $\sigmalos$, but, different from
the data of \citet{Par18}, in our progenitor galaxies the points with
the highest $\alphaIMF$ are not those with the highest $\sigmalos$:
this reflects the fact that the $\sigmalos$ profiles of these models
peak at $R\approx 0.2 \Re$ and slightly decreases towards the centre
(\Fig\ref{fig:rimf}).  By construction, the distribution of the
satellite galaxy in the unequal-mass merger simulations is a
scaled-down version of that of of the main galaxy.

The $\alphaIMF$-$\sigmalos$ distributions of the remnants are
qualitatively similar to those of the progenitor galaxies. In detail,
the distributions tend to be narrower in $\sigmalos$, because in the
radial range $0.1\lesssim R/\Re\lesssim 1$ the remnants' $\sigmalos$
profiles are flatter than those of the progenitors (see upper panel of
\Fig\ref{fig:rimf_norm}).  Moreover, with the only exception of the
$\xistar=0.2$ head-on merger, the range in $\alphaIMF$ spanned by the
remnants tends to be smaller than that spanned by the progenitors, as
a consequence of the mixing of the stellar populations in the central
regions.  The average $\sigmalos$ of the remnant is higher than that
of the main galaxy when $\xistar=1$ and lower when $\xistar=0.2$. When
$\xistar=0.5$ the average $\sigmalos$ is comparable to that of the
main galaxy, with a behaviour that depends in detail on the orbital
angular momentum of the encounter. We note that the remnant of the
head-on $\xistar=0.5$ major merger has a peculiar distribution in the
$\sigmalos\alphaIMF$ plane (squares in the middle panel of
\Fig\ref{fig:sigimf}), which reflects the somewhat unusual $\sigmalos$
and $\alphaIMF$ profiles (squares in \Fig\ref{fig:rimf_xi05}). In this
case the satellite reaches and modifies the central regions of the
main, both inducing mixing in the existing stellar populations and
depositing its own stars. The effect is strongest for intermediate
mass ratios ($\xistar=0.5$), because for higher mass ratios
($\xistar=1$) the progenitors have the same average $\alphaIMF$ and
for lower mass ratios ($\xistar=0.2$) the satellite carries a small
fraction of the total stellar mass of the remnant.  However,
encounters with exactly zero orbital angular momentum are extreme
cases, and we expect the off-axis simulations, which produce more
regular distributions of $\sigmalos$ and $\alphaIMF$, to be more
realistic (see \Sect\ref{sec:set}).

We note that in figure 16 of \citet{Par18}, which is based on a sample
of present-day ETGs, more massive galaxies have higher average
$\sigmalos$, as it is usual. In our minor merger simulations (bottom
panel of \Fig\ref{fig:sigimf}) the remnants have lower average
$\sigmalos$ than their main progenitors, which are naturally less
massive.  This result is not necessarily in contrast with the
observational data, because it must be put in the context of galaxy
evolution (see
\Sects\ref{sec:effectonsigalpha}-\ref{sec:alphasigrel}): we recall
that the observed $\Mstar$-$\sigmae$ relation of ETGs evolves with
redshift in the sense that, at given $\Mstar$, $\sigmae$ tends to be
higher at higher $z$ \citep[see][and references therein]{Can20}.

\begin{figure}
  \centerline{\psfig{file=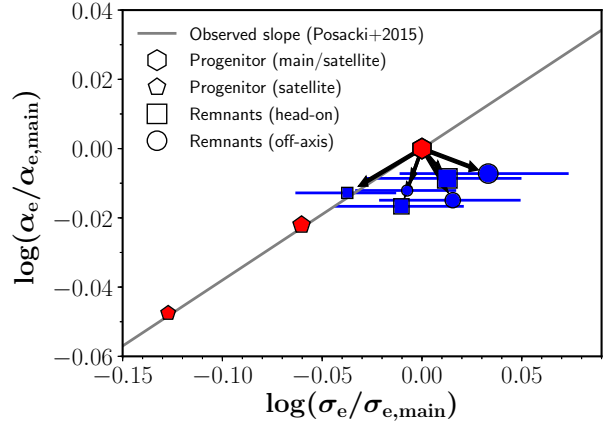,width=\hsize}}
\caption{Effective IMF $\alphae$ (normalized to $\alphae$ of the main
  progenitor galaxy) as a function of effective velocity dispersion
  $\sigmae$ (normalized to $\sigmae$ of the main galaxy) for the
  remnants of simulations 1Dh (head-on, $\xistar=1$), 1Do (off axis,
  $\xistar=1$), 0.5Dh (head-on, $\xistar=0.5$), 0.5Do (off axis,
  $\xistar=0.5$), 0.2Dh (head-on, $\xistar=0.2$) and 0.2Do (off axis,
  $\xistar=0.2$), and of their main (hexagons) and satellite
  (pentagons) progenitor galaxies (in the case $\xistar=1$ the
  satellite progenitor galaxy is identical to the main galaxy, and
  both are represented by the hexagon). The remnants of head-on and
  off-axis merging simulations are represented, respectively, by
  squares and circles, whose size is larger for higher stellar-mass
  ratios $\xistar$. The error bars indicate 1$\sigma$ scatter on the
  effective velocity dispersion due to projection effects on the
  remnants (the $1\sigma$ scatter in effective IMF mismatch parameter
  is comparable to or smaller than the symbol size). The arrows
  connect the position of the main galaxy with those of the remnants
  and thus indicate the transformations produced by the mergers on the
  main galaxy. The grey line indicates the slope of the observational
  $\alphae$-$\sigmae$ relation as determined by \citet{Pos15} for a
  sample of present-day ETGs (\Eq\ref{eq:obsalpsig}).  We note that
  the scatter of the observational relation, $0.12$ dex in $\alphae$
  at given $\sigmae$, is comparable to extent of the vertical axis of
  the plot.}
\label{fig:sigealpha}
\end{figure}

\begin{figure}
  \centerline{\psfig{file=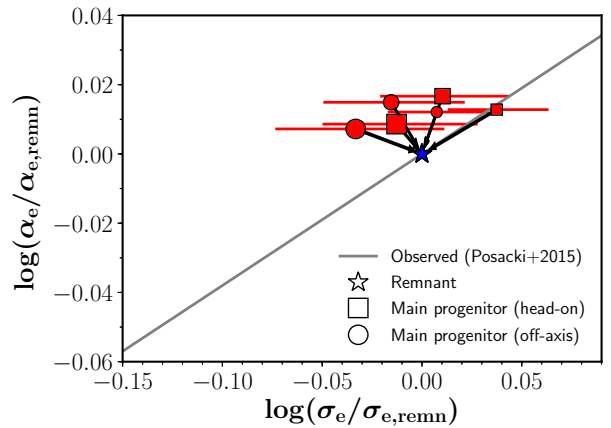,width=\hsize}}
\caption{Same as Fig.\ \ref{fig:sigealpha}, for the same
  simulations, but now $\alphae$ and $\sigmae$ are normalized to
  values $\alphaeremn$ and $\sigmaeremn$ measured for the remnants. By
  definition the remnant, represented by a star, lies at
  $\sigmae/\sigmaeremn=1$ and $\alphae/\alphaeremn=1$.  The
  progenitors are indicated with squares for head-on mergers and
  circles for off-axis mergers, of size increasing with
  $\xistar$. Error bars on the positions of the main progenitors are
  due to $1\sigma$ projection-effect scatter on the remnant's
  effective velocity dispersion.  Here the satellite galaxies are not
  shown. The grey solid line indicates the observed present-day
  $\alphae$-$\sigmae$ relation (\Eq\ref{eq:obsalpsig}). The arrows
  have the same role as in \Fig\ref{fig:sigealpha}.}
\label{fig:sigealpha_back}
\end{figure}

\subsection{Effect of dry mergers on $\sigmae$ and $\alphae$}
\label{sec:effectonsigalpha}

We move here to study the effect of the considered dissipationless
mergers on the global galaxy properties: the effective IMF mismatch
parameter $\alphae$ and the effective velocity dispersion $\sigmae$.
\Fig\ref{fig:sigealpha} shows the behaviour of our simulations in the
$\sigmae\alphae$ plane in which $\alphae$ is normalized to
$\alphaemain$ and $\sigmae$ to $\sigmaemain$. By construction
(i.e.\ as a consequence of our choice of the parameter $\CIMF$), the
main galaxy (hexagon) and the satellite galaxies (pentagons) lie on a
power law $\alphae\propto \sigmae^{0.38}$ with the same slope as the
correlation (\Eq\ref{eq:obsalpsig}) observed for present-day ETGs. The
remnants (squares and circles with error bars) have in all cases
$\alphae$ lower than that of the main progenitor, which is an expected
consequence of the flattening of the $\alphae$ gradient in the
equal-mass mergers and also of the accretion of a lower-$\alphae$
galaxy in the unequal-mass mergers. The final value of $\alphae$
depends more on the mass ratio than on the orbital angular momentum of
the encounter.  The lowest values of $\alphae$ are obtained in the
$\xistar=0.5$ major merger, while the remnants of the $\xistar=0.2$
minor mergers have $\alphae$ intermediate between the $\xistar=1$ and
the $\xistar=0.5$ major mergers. The effect on $\sigmae$ is
variegated, ranging from equal-mass mergers that make $\sigmae$
increase to minor mergers that make $\sigmae$ decrease. With the only
exception of the head-on $\xistar=0.2$ merger (arrow pointing towards
the smallest square in \Fig\ref{fig:sigealpha}), for which the remnant
lies on the same $\alphae$-$\sigmae$ power law followed by the
progenitor galaxies, the effect of the dry merging is to move the
galaxies away from the $\alphae$-$\sigmae$ relation, producing
remnants with low $\alphae$ for their $\sigmae$ (compared to the
progenitor galaxies). The error bars in \Fig\ref{fig:sigealpha} give a
measure of the projection effects in $\sigmae$, due to the fact that
the remnants are not spherically symmetric (the projection effects on
$\alphae$ turn out to be negligible). We note that these projection
effects are well within the intrinsic scatter of the observed
correlation, which is $\approx 0.1$ dex in $\alphae$ at given
$\sigmae$ (see \Sect\ref{sec:intro}) and thus, given the slope of the
relation (\Eq\ref{eq:obsalpsig}), $\approx 0.3$ dex in $\sigmae$ at
given $\alphae$.

In order to compare quantitatively the effect on $\sigmae$ and
$\alphae$ of mergers with different $\xistar$, it is useful to
introduce the quantities
\begin{equation}
 \gammasigma\equiv\frac{\log\sigmaeremn-\log\sigmaemain}{\log\Mstarremn-\log\Mstarmain}=\frac{\log\sigmaeremntilde}{\log(1+\xistar)}
\label{eq:deltasigma}
\end{equation}
and
\begin{equation}
\gammaalpha
  \equiv\frac{\log\alphaeremn-\log\alphaemain}{\log\Mstarremn-\log\Mstarmain}=\frac{\log\alphaeremntilde}{\log(1+\xistar)},
\label{eq:deltaalpha}
\end{equation}  
which measure the variations of, respectively, $\log\sigmae$ and
$\log\alphae$ per unit logarithmic stellar mass increase (the values
of $\sigmaeremntilde$ and $\alphaeremntilde$ are given for each
simulation in \Tab\ref{tab:simu}; we have used
$\Mstarremn/\Mstarmain=1+\xistar$, because the stellar mass loss turns
out to be negligible in the considered mergers). In our set-D
simulations we find $-0.48\lesssim
\gammasigma\lesssim 0.11$ and $-0.16\lesssim
\gammaalpha\lesssim -0.03$, where the lowest
values are for $\xistar=0.2$ and the highest for $\xistar=1$.

For a more direct comparison with observational data, we plot in
\Fig\ref{fig:sigealpha_back} a $\sigmae\alphae$ diagram in which the
remnants are assumed to lie on the observed present-day
$\alphae$-$\sigmae$ relation (solid line). Note that, different from
\Fig\ref{fig:sigealpha}, in \Fig\ref{fig:sigealpha_back} we normalize
$\alphae$ to $\alphaeremn$ and $\sigmae$ to $\sigmaeremn$, so, by
construction, all the remnants are at the same point, which is assumed
to lie on the observed present-day $\alphae$-$\sigmae$ relation.  In
\Fig\ref{fig:sigealpha_back} the circles and squares indicate the
positions of the main galaxies of the simulations, assuming for
$\sigmaeremn$ and $\alphaeremn$ the mean values measured over all the
considered projections of the corresponding remnants, and the error
bars, as in \Fig\ref{fig:sigealpha}, give a measure of the projection
effects due to deviations from spherical symmetry.  In
\Fig\ref{fig:sigealpha_back} the squares and the circles can be
considered the progenitors of present-day ETGs that lie on the
observed $\alphae$-$\sigmae$ relation. All the progenitors, but that
of the head-on $\xistar=0.2$ merger, lie above the $\alphae$-$\sigmae$
relation.

\subsection{Merger-driven evolution of the $\alphae$-$\sigmae$ relation}
\label{sec:alphasigrel}

We recall that so far we have not considered the effects of full
merging hierarchies, but only the effects of single binary mergers.
Here we attempt to predict the cosmological evolution of the
$\alphae$-$\sigmae$ relation based on the results of our simulations.

\subsubsection{Purely dry merging hierarchies}
\label{sec:hier}

When cosmologically motivated merging hierarchies are considered,
massive galaxies in the redshift range $0\lesssim z\lesssim 2$
experience merging histories with average mass-weighted merger stellar
mass ratio $\avxistar$ in the range $0.3\lesssim \avxistar \lesssim
0.5$ (\citealt{Son17}).  We thus expect that in a cosmologically
motivated merging history a massive ETG moves in the $\sigmae\alphae$
plane along a direction which is in between those of the $\xistar=0.2$
and $\xistar=0.5$ simulations (arrows starting from small and
intermediate circles and squares in \Fig\ref{fig:sigealpha_back}).

\begin{figure}
  \centerline{\psfig{file=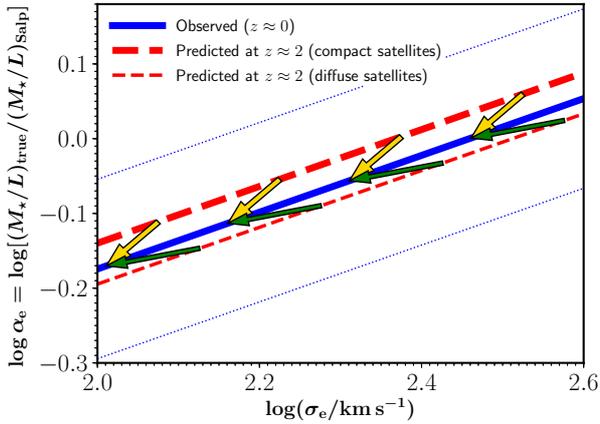,width=\hsize}}
  \caption{Toy model, based on the results of the $N$-body simulations
    presented in this work, representing the dry-merger driven
    evolution of the $\alphae$-$\sigmae$ relation of ETGs. Here
    $\sigmae$ is expressed in physical units and $\alphae$ is defined
    as $(\Mstar/L)_{\rm true}/(\Mstar/L)_{\rm Salp}$, thus taking the
    Salpeter IMF as reference. The solid line represents the best fit
    found by \citet{Pos15} for present-day ETGs
    (\Eq\ref{eq:obsalpsig}) and the dotted lines indicate the
    intrinsic scatter of the observed correlation.  The thick arrows
    indicate the evolution of individual galaxies expected from
    $z\approx 2$ progenitors to $z\approx 0$ descendants (assuming
    that, on average, the stellar mass of individual galaxies has
    grown by a factor of three), when the accreted satellites are
    compact, such as those considered in our simulations.  The thick
    dashed line represents the correlation predicted in this case at
    $z\approx 2$. The thin arrows indicate qualitatively how the
    corresponding thick arrows are modified if the accreted satellites
    are so diffuse to lead to an opposite evolution of the correlation
    (thin dashed line).}
\label{fig:sigealpha_toy}
\end{figure}

Given that in our model the satellite and main galaxies lie on a
$\alphae$-$\sigmae$ power law with the same slope as that observed at
$z\approx 0$, our results suggest a possible scenario in which the
$\alphae$-$\sigmae$ relation evolves by maintaining its slope and
changing its normalisation: at given $\sigmae$, $\alphae$ is higher at
higher redshift. Such a scenario is qualitatively represented in
\Fig\ref{fig:sigealpha_toy}. While $\alphae$ invariably decreases in
this model, the evolution of $\sigmae$ is more uncertain: even
limiting to simulations with $\xistar=0.2$ and $\xistar=0.5$,
bracketing the cosmologically motivated value of $\avxistar$,
$\sigmae$ decreases in some cases and increases in others.

Let us consider an individual galaxy that experiences a merging
hierarchy that produces an increase in stellar mass
$\Delta\log\Mstar$: the corresponding variations in $\sigmae$ and
$\alphae$, are given by
$\Delta\log\sigmae=\Delta\log\Mstar\av{\gammasigma}$ and
$\Delta\log\alphae=\Delta\log\Mstar\av{\gammaalpha}$, where
$\av{\cdots}$ indicates the average over the merging history.  Given
the expected values of $\avxistar$, we can build a toy model by
estimating these averages as the mean values obtained for our four
set-D simulations with either $\xistar=0.5$ or $\xistar=0.2$:
$\av{\gammasigma}=-0.135$ and $\av{\gammaalpha}=-0.123$.  The thick
yellow arrows in \Fig\ref{fig:sigealpha_toy} indicate the effect of
these variations for individual galaxies increasing their stellar
masses by a factor of three as a consequence of dry mergers and ending
up onto the mean observed $\alphae$-$\sigmae$ relation at $z\approx0$.
Given that a factor of three increase in stellar mass is expected from
$z\approx 2 $ to $z\approx 0$ (e.g.\ \citealt{Son17}), we can
interpret the starting points of the arrows as $z\approx 2$.  In the
hypothesis that ETGs over the entire $\sigmae$ range have experienced
similar accretion histories, the net effect of the evolution of
individual galaxies is that at higher redshift the $\alphae$-$\sigmae$
had similar slope but higher normalisation (at given $\sigmae$,
$\alphae$ is predicted to be higher at higher $z$; solid and dashed
thick lines in \Fig\ref{fig:sigealpha_toy}). We note that, though the
variation in stellar mass is as high as a factor of three, the
predicted variation in the normalisation of the $\alphae$-$\sigmae$
relation is just $0.04$ dex, much smaller than the intrinsic scatter
of the present-day relation ($0.12$ dex; thin dotted lines in
\Fig\ref{fig:sigealpha_toy}).

We recall that in our simulations the satellite and main galaxies are
assumed to follow $\Re$-$\Mstar$ and $\sigmae$-$\Mstar$ relations with
slopes similar to those observed for massive ETGs
(\Sect\ref{sec:progenitor}): in this sense the satellites can be
considered {\em compact}, following the terminology of
\citet{Hil++13}. This choice is consistent with the fact that in the
present-day Universe the $\Re$-$\Mstar$ and $\sigmae$-$\Mstar$
relations are found to be similar for central and satellite galaxies
\citep{Spi17,Wan20}.  However, the properties of satellite galaxies at
higher $z$ are poorly constrained, so it is not excluded that the
accreted satellites can be {\em diffuse}, i.e.\ having lower $\sigmae$
and larger $\Re$ for their stellar mass \citep{Hil++13}. The
assumption of compact satellites might be observationally motivated by
the fact that there is no evidence of evolution of the slopes of the
$\Re$-$\Mstar$ and $\sigmae$-$\Mstar$ relations \citep{vdWel14,Can20},
though a steepening of the $\sigmae$-$\Mstar$ relation with increasing
redshift is not excluded \citep{Can20}, which might instead favour the
hypothesis of diffuse satellites.

Dissipationless mergers with diffuse satellites have been studied with
$N$-body simulations by \citet{Hil++13}. Based on the results of Hilz
et al.'s simulations and on well-known properties of interacting
stellar systems, it is easy to predict qualitatively how the
dry-merger driven evolution of galaxies in the $\sigmae\alphae$ plane
changes if the accreted satellites are diffuse. A diffuse satellite,
being loosely bound, is easily disrupted during the merger and
deposits most of its stars in the outskirts of the main galaxy, thus
producing a remnant with central $\alphaIMF$ similar to the progenitor
and lower $\alphaIMF$ in the outskirts (thus steepening the original
$\alphaIMF$ gradient).  As a consequence, $\alphae$ of the remnant
will be only slightly lower than that of the main progenitor.  Because
of the accretion of loosely bound stars, the velocity dispersion of
the remnant is lower than in the case of compact satellites
\citep{NJO09}, so $\sigmae$ decreases more when the satellites are
diffuse. Qualitatively, the net effect is that dry mergers with
diffuse satellites, compared to those with compact satellites, move
galaxies more horizontally in the $\sigmae\alphae$ plane (thin green
arrows in \Fig\ref{fig:sigealpha_toy}), possibly leading to a
different evolution of the $\alphae$-$\sigmae$ relation, with lower
$\alphae$ at higher $z$, at given $\sigmae$ (thin dashed line in
\ref{fig:sigealpha_toy}).

\subsubsection{The effect of dissipation and star formation}

The simulations used in this work are admittedly idealized, not only
because they are not within a fully cosmological context, but also
because they are completely dissipationless. Present-day ETGs are poor
in cold gas and have a stellar component that is dominated by old
stellar populations, so if they experienced mergers in relatively
recent times, these mergers must have been essentially dry. Detailed
analyses of the stellar population properties of ETGs indicate that at
most a few per cent of their stellar mass formed at $z\lesssim 1$
(e.g.\ \citealt{Tra++00a}, \citealt{Tho++10}; see also \citealt{Son14}
and references therein). Though this fraction is small, such star
formation could affect non-negligibly the evolution of both $\sigmae$
and $\alphae$ if, as expected, it occurs in the central regions of the
galaxies. Moreover, star formation might have contributed more at
$z\gtrsim1$.

Compared to purely dry mergers, slightly `wet' mergers, i.e.\ with
some dissipation and star formation, are expected to produce remnants
more compact and thus with higher stellar velocity dispersion
\citep{Rob++06c,Cio07,Son14}. The effect of dissipation and star
formation on $\alphae$ depends on the IMF of the stars that are formed
in the star formation episodes occurring during the mergers, which is
of course highly uncertain. Based on the proposal that the IMF is
heavier when the pressure of the star-forming gas is higher
(e.g.\ \citealt{Bar18} and references therein), one might expect that
at lower $z$ stars form with IMF lighter than that of stars formed at
higher redshift (because the pressure of the gas form which stars form
tends to decrease with cosmic time; \citealt{Bar19b}).  However, as
far as we know, it is not excluded, either theoretically or
observationally, that at lower $z$ stars can form with heavier IMF.

\subsubsection{Comparison with observations and with previous models}
\label{sec:comp}

The theoretical predictions on the evolution of the
$\alphae$-$\sigmae$ relation can be tested with measurements of
$\alphae$ and $\sigmae$ at different redshifts. However, so far such
measurements are relatively rare and it is difficult to draw robust
conclusions. On the one hand, \citet{Son15} find that in an observed
sample of lens ETGs, $\alphae$ tends to decrease with increasing
redshift, at fixed $\sigmae$, out to $z\approx 0.8$. On the other
hand, \citet{Mar15b} find that massive ETGs at $z\approx 1$ have IMF
similar to (or slightly heavier than) present-day ETGs with comparable
stellar velocity dispersion. Also \citet{She14} find that massive ETGs
at $z\approx 0.75$ have, at given $\sigmae$, IMF similar to that of
lower-$z$ galaxies (or slightly heavier; see
\citealt{Son17}). Recently, \citet{Men20} measured $\alphae$ and
$\sigmae$ for a sample of quiescent galaxies at $1.4<z<2.1$, finding a
steeper $\alphae$-$\sigmae$ relation, which overlaps with the
$z\approx 0$ relation at the high-$\sigmae$ end.  The trend found by
\citet{Son15} and \citet{Men20} (higher-$z$ galaxies tend to have
lower $\alphae$ at given $\sigmae$) appears in tension with the
compact-satellite evolutionary model depicted in
\Fig\ref{fig:sigealpha_toy} and more consistent with the hypothesis of
diffuse satellites. It should be noted, though, that the measurements
of both \citet{Son15} and \citet{Men20} are relying on a set of
assumptions, most notably that of a spatially constant stellar
mass-to-light ratio at all redshifts. This assumption can have a big
impact on their estimates of the IMF (see e.g.\ \citealt{Son18} and
\citealt{Ber18}) and, consequently, on their measured trend with
redshift.

The theoretical model of \citet{Son17}, who found a negligible
evolution of the $\alphae$-$\sigmae$ relation (i.e.\ $\alphae$ at
given $\sigmae$ independent of redshift), lies in between the
compact-satellite and the diffuse-satellite scenarios depicted in
\Fig\ref{fig:sigealpha_toy}.  In our simulations both the effective
velocity dispersion and the effective IMF mismatch parameter of the
remnants are computed self-consistently accounting for the internal
kinematics, structure and $\alphaIMF$ distribution of the $N$-body
systems.  In their statistical approach, \citet{Son17} assume that
$\sigmae$ is proportional to the virial velocity dispersion and that
$\alphae$ of the remnant is the weighted mean of $\alphae$ of the
progenitor galaxies. When this weighted mean is adopted, $\alphae$ of
the remnant is overestimated compared to the case of compact
satellites, in which most of the variation of $\alphaIMF$ occurs in
the center, so, in this respect, the model of \citet{Son17} is closer
to the diffuse-satellite scenario.  Qualitatively, our
compact-satellite scenario predicts an evolution of the
$\alphae$-$\sigmae$ relation broadly consistent with the cosmological
model `LoM' of \citet{Bar19b} and with that of \citet{Bla17}, which
both predict higher $\alphae$ at higher redshift, for given $\sigmae$.

\section{Conclusions}\label{sec:concl}  
 
We have studied the effect of dissipationless (dry) mergers on the
distribution of the IMF mismatch parameter $\alphaIMF$ in ETGs using
the results of dissipationless binary major and minor merging
simulations. Our main conclusions are the following.
\begin{itemize}
\item Dissipationless mergers tend to make the $\alphaIMF$ profiles of
  ETGs shallower, and in particular to produce flat central
  ($R\lesssim 0.3 \Re$) $\alphaIMF$ distributions. 
\item Dissipationless mergers do not alter significantly the shape of
  the spatially resolved distributions in the $\sigmalos\alphaIMF$
  space: when the progenitor galaxies have realistic distributions in this
  space, this is true also for the merger remnants.
\item Individual galaxies undergoing dry mergers move, in the space of
  integrated quantities $\alphae\sigmae$, by decreasing their
  $\alphae$, due to the erosion of $\alphaIMF$ gradients and mixing
  with stellar populations with lighter IMF, while their $\sigmae$ can
  either decrease or increase, depending on the merging orbital
  parameters and mass ratio. $\sigmae$ tends to decrease in
  cosmologically motivated merging histories.
\item The dry-merger driven evolution of the $\alphae$-$\sigmae$
  relation of ETGs depends on the nature of the accreted satellites:
  galaxies of given $\sigmae$ are expected to have higher $\alphae$ at
  higher redshift if the satellites are compact, but the trend can be
  opposite if the satellites are sufficiently diffuse.
\end{itemize}  

The effects of dry mergers on the $\alphaIMF$ distribution and on the
$\alphae$-$\sigmae$ predicted by our model are not dramatic and are
thus broadly consistent with the currently available observational
constraints, which are however limited and somewhat controversial.
Some observational estimates of the evolution of the
$\alphae$-$\sigmae$ relation of ETGs \citep{Son15,Men20} indicate that
$\alphae$ at given $\sigmae$ tends to be lower than at higher
$z$. This is in tension with the predictions of dry-merger simulations
in which the satellites are compact, and suggest that accretion of
diffuse satellites might be invoked to reconcile a dry-merging driven
evolution with observational data. Additional measurements of the
properties of the IMF of ETGs beyond the present-day Universe are
necessary to further test the two-phase formation model of massive
ETGs, in which essentially dissipationless mergers have an important
role at $z\lesssim 2$.  A very promising possibility is to estimate
$\alphae$ of lens ETGs by combining constraints on the total mass from
gravitational lensing with spatially resolved kinematics
\citep[see][]{Tre10}, which will be feasible over a significant
redshift range with forthcoming telescopes and instruments
\citep[e.g.][]{Sha18}.

\section*{Data Availability}

The data underlying this article will be shared on reasonable request to the corresponding author.

\section*{Acknowledgements}

FC acknowledges support from grant PRIN MIUR 20173ML3WW$\_$001.

\bibliography{biblio_imf}
\bibliographystyle{mnras}

\appendix

\section{Varying the $\alphaIMF$ profiles of the progenitor galaxies}
\label{sec:varimf}

\begin{figure} 
  \centerline{\psfig{file=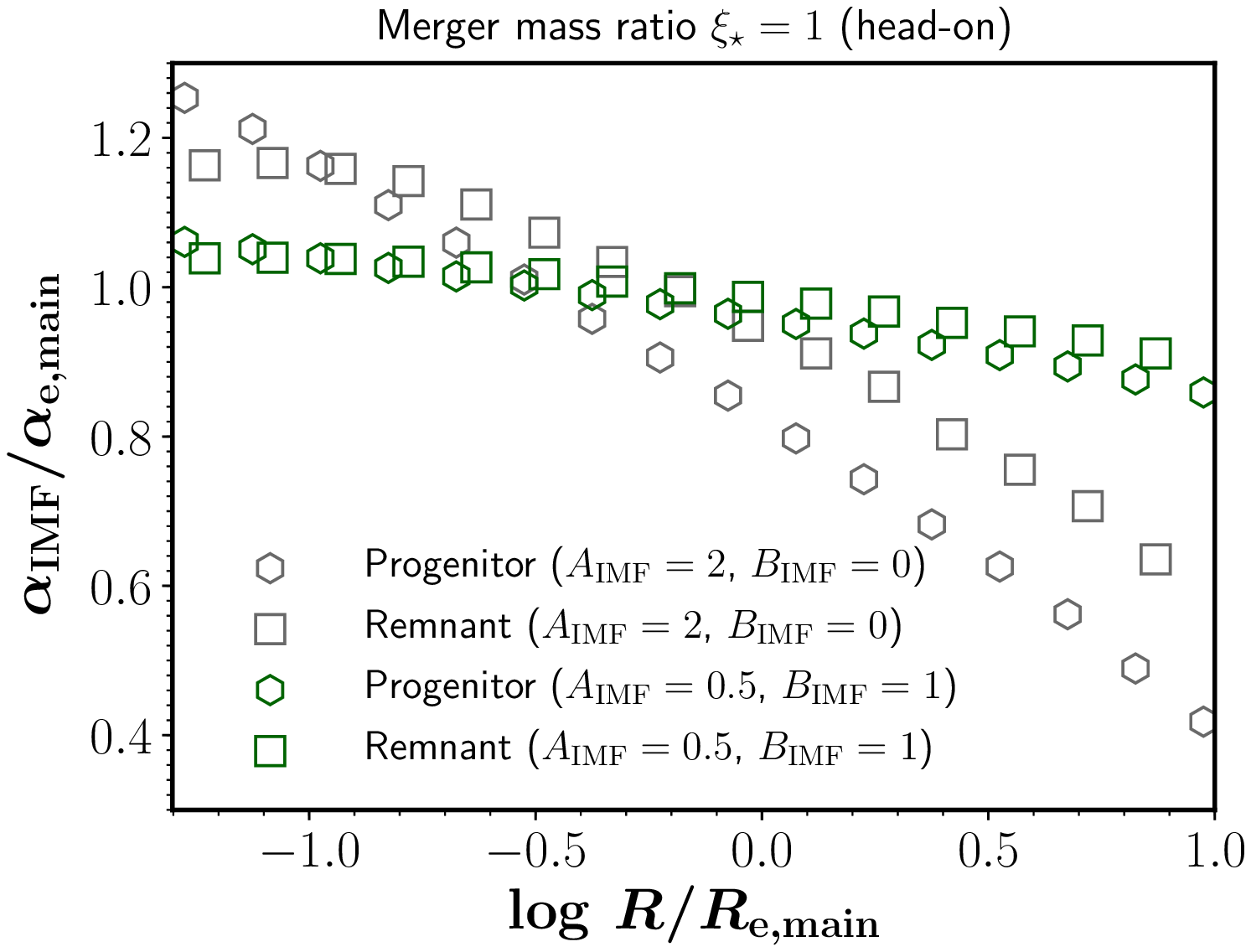,width=\hsize}}
  \centerline{\psfig{file=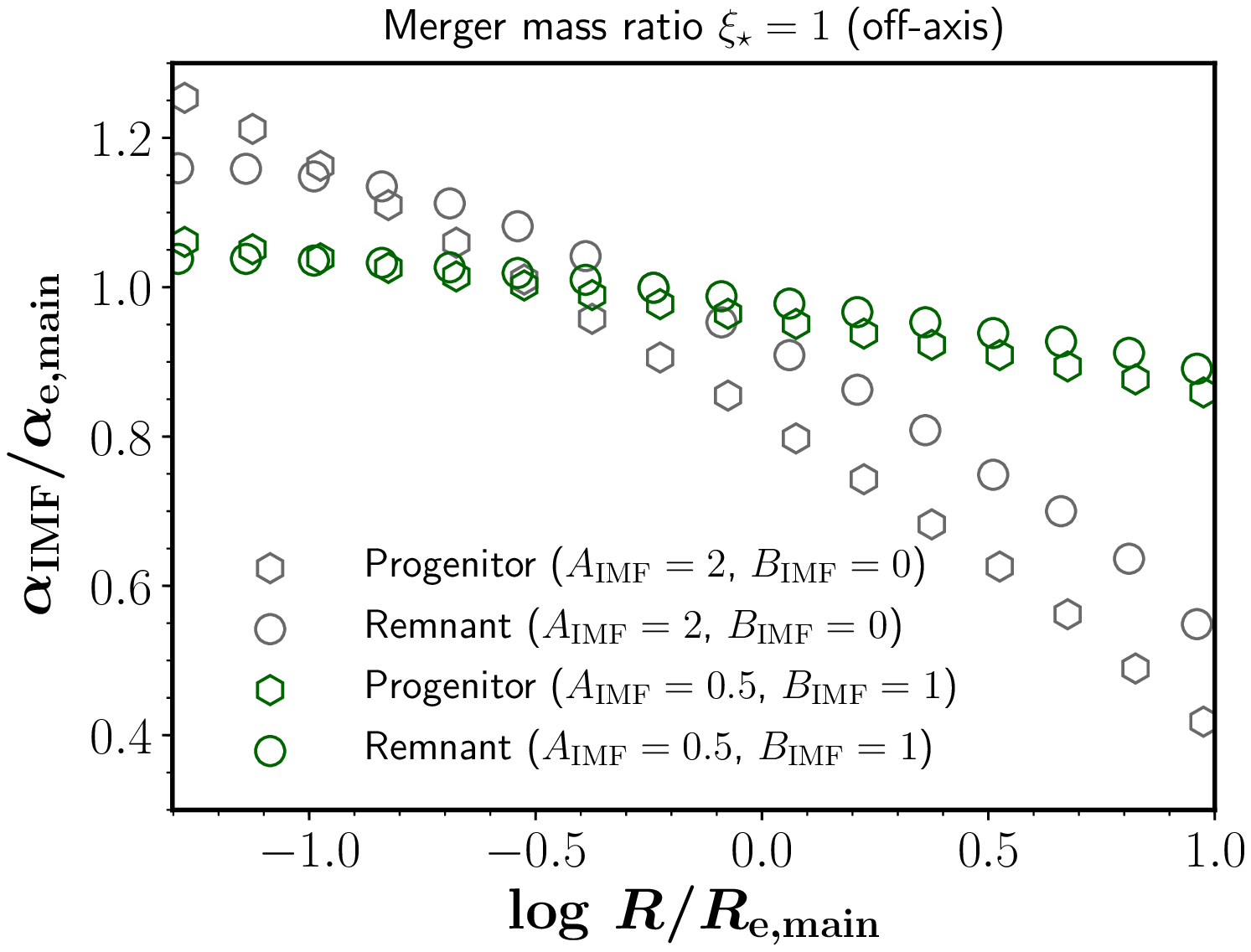,width=\hsize}}
 \caption{Same as lower panel of Fig.\ \ref{fig:rimf_xi1}, but for the
   equal-mass head-on merger models D1h and D1h\_bis (upper panel),
   and off-axis merger models D1o and D1o\_bis (lower panel).  Models
   D1h\_bis and D1o\_bis are based on the same simulations as models
   D1h and D1o, respectively, but in post-processing different values
   of $\AIMF$ and $\BIMF$ are assumed. The progenitor galaxies
   (hexagons) and the remnants (squares and circles) of models D1h and
   D1o ($\AIMF=2$, $\BIMF=0$) are represented by grey symbols, while
   those of models D1h\_bis and D1o\_bis ($\AIMF=0.5$, $\BIMF=1$) by
   green symbols.}
\label{fig:rimf_xi1_mod}
\end{figure}

\begin{figure} 
  \centerline{\psfig{file=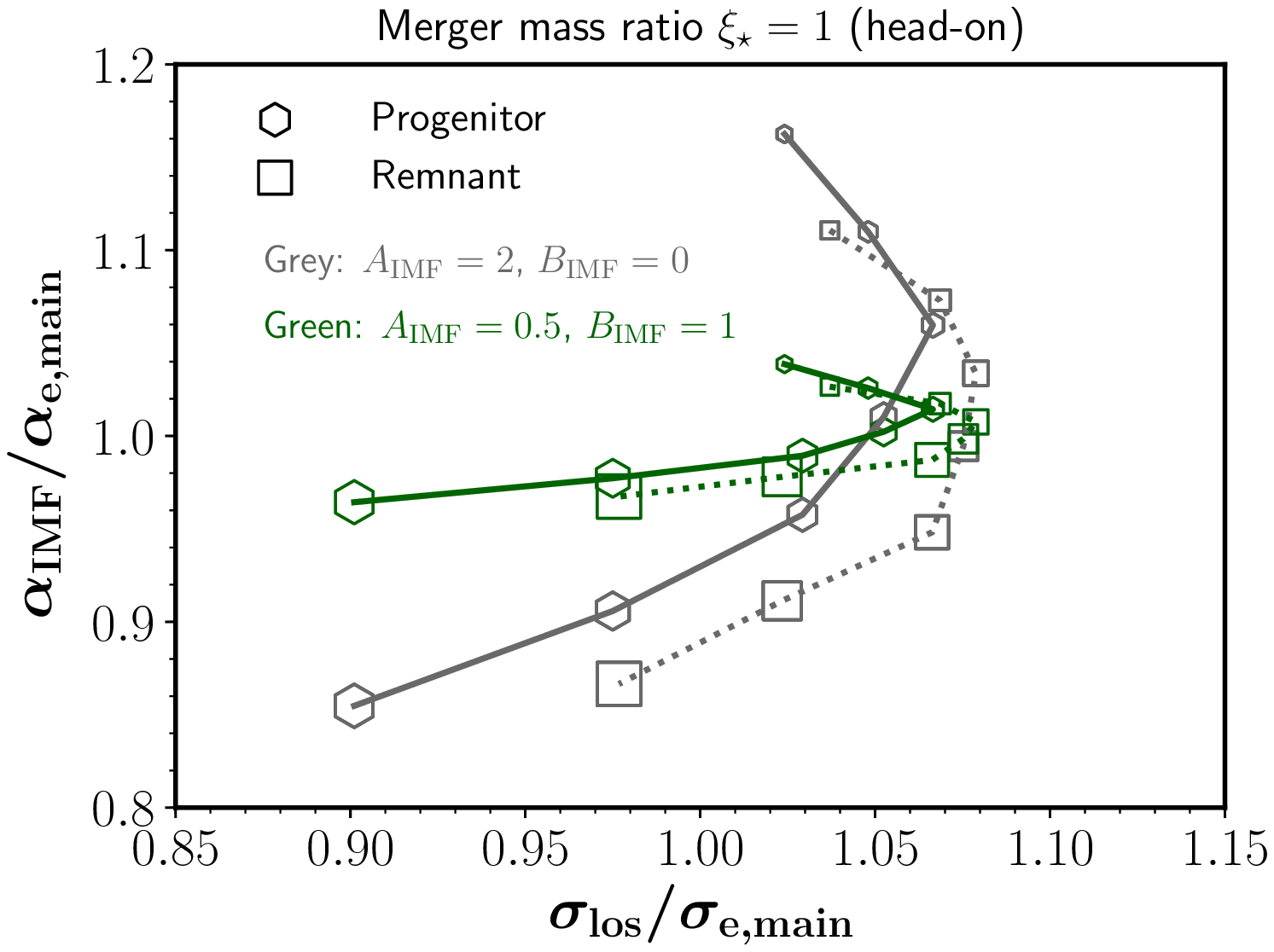,width=\hsize}}
  \centerline{\psfig{file=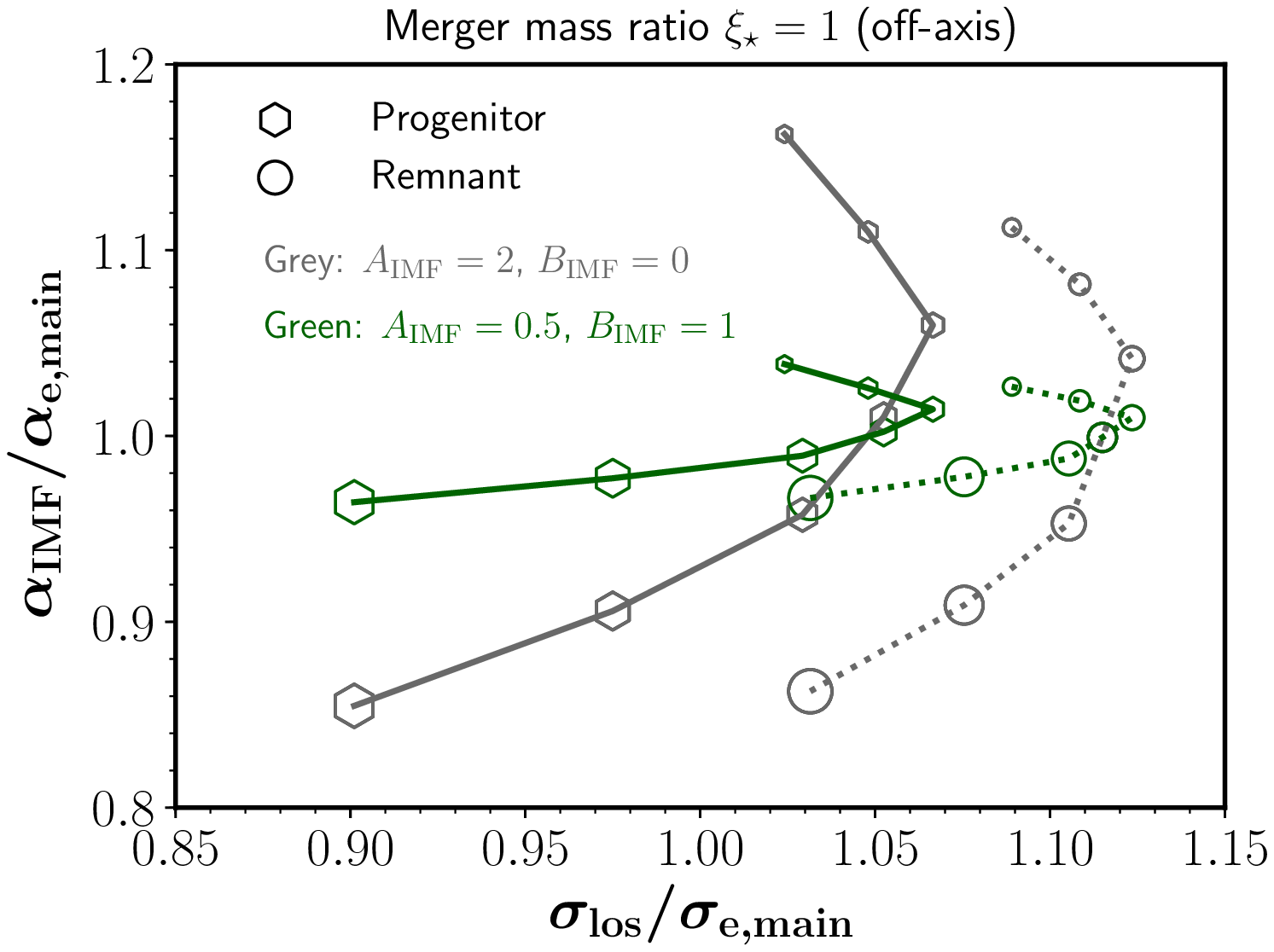,width=\hsize}}
 \caption{Same as top panel of Fig.\ \ref{fig:sigimf}, but for the
   equal-mass head-on merger models D1h and D1h\_bis (upper panel),
   and off-axis merger models D1o and D1o\_bis (lower panel).  Symbols
   and colours are the same as in \Fig\ref{fig:rimf_xi1_mod}.}
\label{fig:sigimf_xi1_mod}
\end{figure}

\begin{figure}
  \centerline{\psfig{file=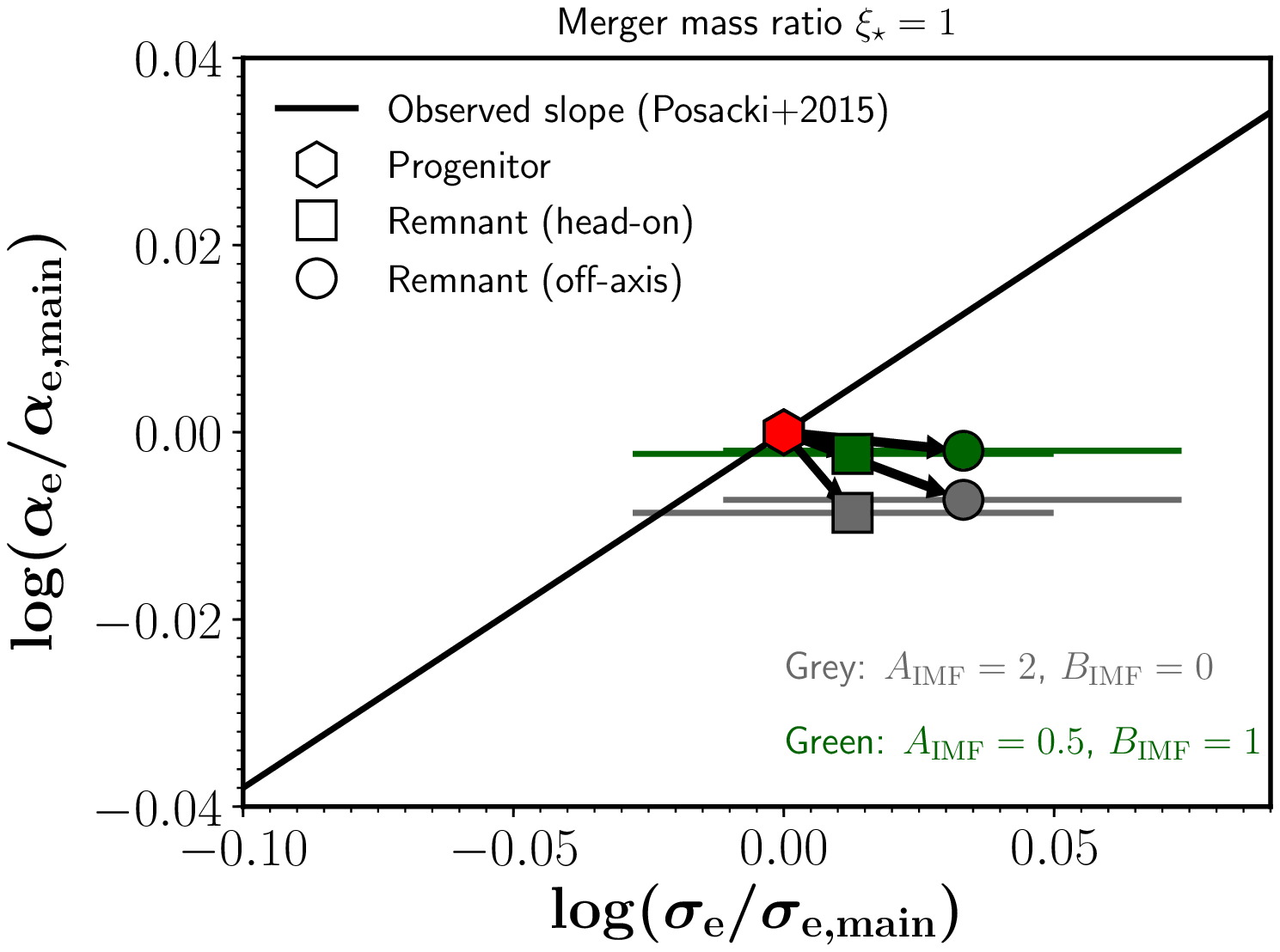,width=\hsize}}
  \caption{Same as Fig.\ \ref{fig:sigealpha}, but for progenitor galaxy
    (hexagon) and the remnants of the equal-mass merger models D1h
    (grey square), D1o (grey circle), D1h\_bis (green square) and
    D1o\_bis (green circle).}
\label{fig:sigealpha_xi1_mod}
\end{figure}

In \Sect\ref{sec:res} we have described how we assign $\alphaIMF$ to
the stellar particles of the simulation by choosing the values of the
dimensionless parameters $\AIMF$, $\BIMF$ and $\CIMF$.  Here we
illustrate the effect of changing the values of these parameters. We
focus on equal-mass mergers simulations and we fix in all cases
$\CIMF=1$ so that the two progenitor galaxies are identical, not only
structurally and kinematically, but also in terms of distribution of
$\alphaIMF$. In particular, we describe here the properties of models
D1h\_bis and D1o\_bis, comparing them with those of models D1h and
D1o. Models D1h\_bis and D1o\_bis are based on the same simulations as
models D1h and D1o, respectively, but in post-processing different
values of $\AIMF$ and $\BIMF$ are assumed (see
\Tab\ref{tab:simu}). The progenitor galaxies of models D1h\_bis and
D1o\_bis have weaker dependence of $\alphaIMF$ on the integral of
motion $Q$ ($\AIMF=0.5$ and $\BIMF=1$) and thus shallower gradients of
$\alphaIMF$ (\Fig\ref{fig:rimf_xi1_mod}) than those of models D1h and
D1o, which have $\AIMF=2$ and $\BIMF=0$. In this sense, models
D1h\_bis and D1o\_bis are less realistic than models D1h and D1o, when
compared to the $\alphaIMF$ gradients inferred for observed ETGs (see
\Sect\ref{sec:imfprofiles}). Due to the shallow $\alphaIMF$ profiles of
their progenitor galaxies, the remnants of models D1h\_bis and
D1o\_bis have normalized $\alphaIMF$ profiles almost indistinguishable
from those of the progenitors
(\Fig\ref{fig:rimf_xi1_mod}). \Fig\ref{fig:sigimf_xi1_mod} shows that
both the progenitors and the remnants of models D1h\_bis and D1o\_bis
have distributions significantly different from the progenitors and
the remnants of models D1h and D1o.  For given $\AIMF$ and $\BIMF$,
the distributions in the $\sigmalos\alphaIMF$ space of the remnants
and the progenitors differ essentially only for variations in
$\sigmalos$.  Consistently, also the merger-driven evolution in the in
the $\sigmae\alphae$ space is weaker for models D1h\_bis and
D1o\_bis than for models D1h and D1o
(\Fig\ref{fig:sigealpha_xi1_mod}). Models D1h\_bis and D1o\_bis are
presented here only for the purpose of illustrating the effect of
changing the values of $\AIMF$ and $\BIMF$, but, for the aforementioned
reasons, should not be considered as representative of real ETGs as
models D1h and D1o.

\section{Mergers between non-homologous galaxies}
\label{sec:varstruct}

\begin{figure} 
  \centerline{\psfig{file=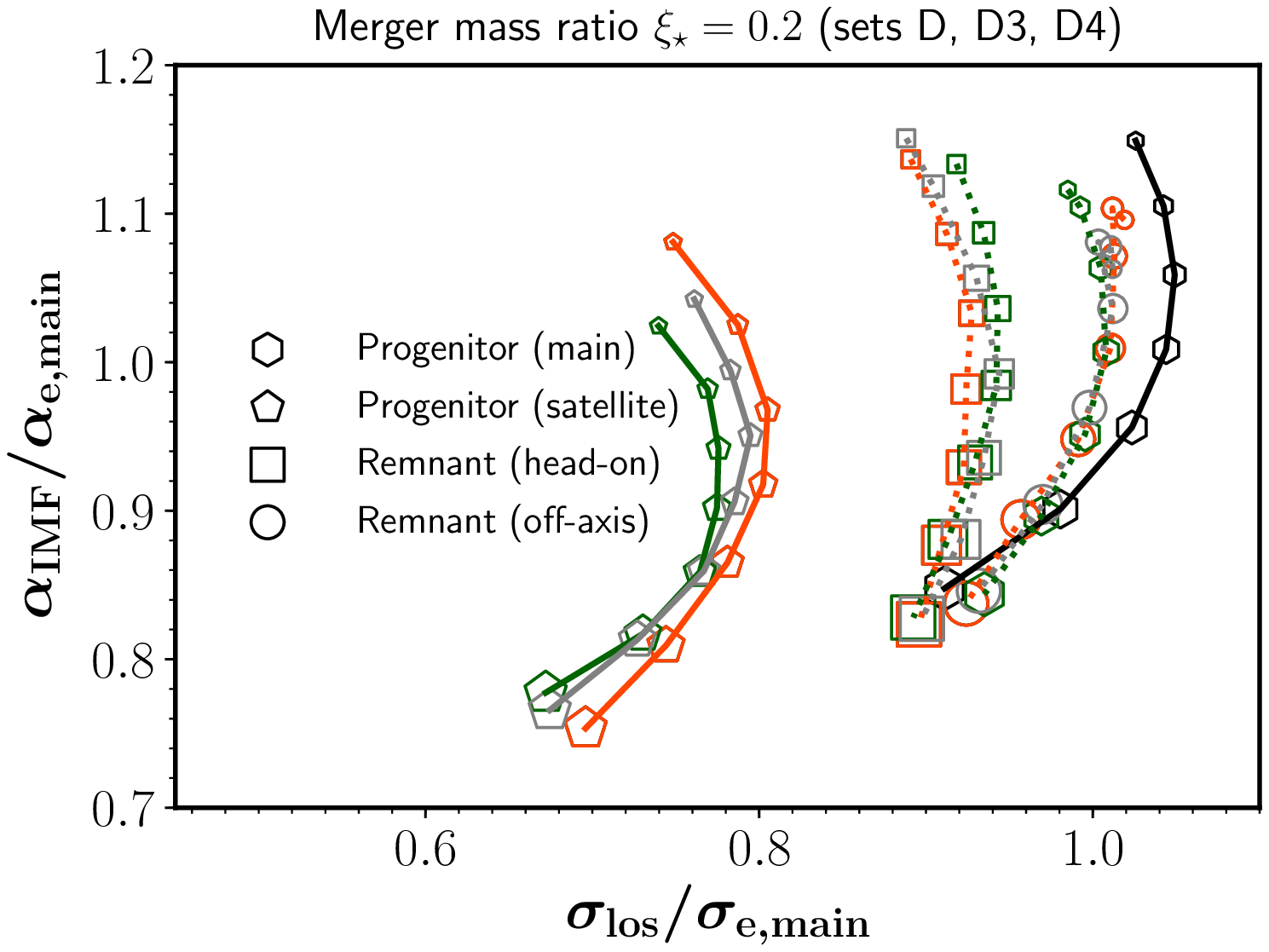,width=\hsize}}
  \caption{Same as bottom panel of Fig.\ \ref{fig:sigimf}, but for the
    $\xistar=0.2$ head-on merger simulations 0.2Dh (set D; grey),
    0.2D3h (set D3; red) and 0.2D4h (set D4; green), and off-axis
    merger simulations 0.2Do (set D; grey), 0.2D3o (set D3; red) and
    0.2D4o (set D4; green). }
\label{fig:sigimf_xi02_D34}
\end{figure}

In \Sect\ref{sec:res}, we have presented results of simulations (set
D) in which the main and satellite galaxies are homologous stellar
systems. Of course, this simplification is not fully justified,
especially in the case of minor mergers, because we expect that
galaxies of different stellar mass differ in their structural
properties, for instance in the distribution and amount of dark
matter, relative to the baryonic matter \citep{Wec18}. Here we
consider $\xistar=0.2$ simulations of sets D3 (0.2D3h and 0.2D3o) and
D4 (0.2D4h and 0.2D4o), in which the structure of the satellite
progenitor galaxy differs significantly from that of the main galaxy,
and we compare them with the corresponding simulations of set D (0.2Dh
and 0.2Do).  The values of the parameters of these sets of simulations
is reported in \Tab\ref{tab:simu} and are chosen to span the range of
values expected for real galaxies, with dark-to-luminous mass ratios
ranging from $35$ to $70$ (see \citealt{Son14}).  Due to the
non-homology, also the $\sigmalos$ profiles of the satellite
progenitor galaxies are not identical in simulations of sets D, D3 and
D4, but the $\sigmalos$ profiles of the remnants are almost
indistinguishable out to $\approx3\Re$.  The values of $\CIMF$ are
such that the satellite and main galaxies lie on the same
$\alphae$-$\sigmae$ power-law relation (\Eq\ref{eq:obsalpsig}), and
are thus different for the three sets of $\xistar=0.2$ simulations,
but the $\alphaIMF$ profiles of the remnants are similar in the three
cases.  Here we show only the distributions of $\alphaIMF$ as a
function of $\sigmalos$ (\Fig\ref{fig:sigimf_xi02_D34}), which for the
remnants of simulations of sets D3 and D4 are very similar to those of
the corresponding simulations of set D. We thus conclude that our
results do not depend significantly on the amount and distribution of
dark-matter in the satellite galaxy.

\end{document}